\documentclass[11pt,a4paper]{article}
\usepackage{jcappub}
\usepackage{multirow}
\usepackage{graphicx}
\usepackage{amsmath,amsfonts,amsthm}
\usepackage{array}
\usepackage{tcolorbox}

\usepackage[export]{adjustbox}

\usepackage{siunitx}

\usepackage{mdframed}
\definecolor{light-gray}{gray}{0.98}

\usepackage[normalem]{ulem}
\usepackage{newcommands}

\DeclareGraphicsExtensions{.png,.pdf}
\graphicspath{{./figs/}{./}}

\subheader{LAPTH-021/26, LUPM:26-001}

\title{In-depth analysis of the clustering of dark matter particles around primordial black holes. Part III: CMB constraints}
\author[a]{Julien Lavalle,}
\author[a]{Vivian Poulin,}
\author[b]{and Pierre Salati}
\affiliation[a]{Laboratoire Univers et Particules de Montpellier (LUPM), Universit\'e de Montpellier \& CNRS, Place Eug\`ene Bataillon, 34095 Montpellier Cedex 05, France}
\affiliation[b]{Universit\'e Grenoble Alpes, USMB, CNRS, LAPTh, 9 chemin de Bellevue, Annecy-le-Vieux, F-74941 Annecy, France}
\emailAdd{lavalle@in2p3.fr}
\emailAdd{vivian.poulin@umontpellier.fr}
\emailAdd{pierre.salati@lapth.cnrs.fr}

\abstract{In a mixed dark matter scenario in which primordial black holes (PBHs) would co-exist with thermally produced self-annihilating particles, one expects the former to be surrounded by extremely dense halos made of the latter, built up during radiation domination. Here, as a continuation of previous work, we derive observational limits on such a scenario from a full statistical analysis of cosmic microwave background (CMB) data. We quantify how a tiny fraction $\fbh$ of PBHs could restrict the parameter space available to thermal particle dark matter, limiting the $s$-wave annihilation cross section to values $\lesssim 10^{-30}\,{\rm cm^3/s}\,(\mchi/100\,{\rm GeV})\,(\fbh/10^{-6})^{-3}$ if PBHs are typically heavier than $\sim 10^{-10}\,\Msun$, which can also be turned into constraints on PBHs in this mass range. In contrast, asteroid mass or lighter PBHs could live in perfect peace with these particles. Finally, we shortly discuss the implications of the recent tentative interpretation of Subaru-HSC microlensing events as PBHs.}

\arxivnumber{2604.xxxxx}

\begin{document}
\maketitle

\section{Introduction}
\label{sec:intro}
The origin of dark matter (DM) has been very actively tracked down for decades \cite{Peebles1982a,JungmanEtAl1996,Feng2010,CirelliEtAl2024}, but has remained elusive so far. So far, Occam's razor has certainly pushed the community to concentrate on single candidates one by one. However, mischievously enough, the universe could actually be populated with several DM species. On the one hand, particle DM, be it thermal or not (\eg~axionic, etc.), finds strong motivations in theoretical particle physics \cite{Feng2010,CirelliEtAl2024}, and current searches have started to probe the relevant parameter space only recently (for a bit more than a decade) owing to continuous and impressive experimental developments over a much longer run. On the other hand, the recent advent of gravitational wave (GW) detectors has prompted a renewed attention on primordial black hole (PBH) DM \cite{ZeldovichEtAl1967,Hawking1971,CarrEtAl1974,Chapline1975,BirdEtAl2016,CarrEtAl2023}, which, while representing an appealing alternative DM candidate essentially independent from particle physics beyond the standard model, had previously lost interest because of strong observational constraints. These include negative results from microlensing searches \cite{AlcockEtAl2000,TisserandEtAl2007,Green2017,PetacEtAl2022}, which is well suited to target the QCD-transition mass range \cite{Jedamzik1997}, and even stronger constraints at higher masses extracted from the cosmic microwave background (CMB) radiation data \cite{RicottiEtAl2008a,Ali-HaiemoudEtAl2017,PoulinEtAl2017}. An additional issue with PBHs relates to the fact that designing a natural formation scenario in the framework of standard inflation remains challenging \cite{BallesterosEtAl2018a,DeLucaEtAl2021a,ColeEtAl2023}. Still, in spite of significant constraints over a very extended mass range also coming from GWs themselves (see more general reviews in \eg~\cite{CarrEtAl2020b,GreenEtAl2021a,CarrEtAl2021d}), PBHs could still make up a fraction of the DM (if not all \cite{DeLucaEtAl2021a,CarrEtAl2023}) and then stand as a DM subcomponent. As a matter of fact, a single and unambiguous detection of a subsolar mass merger event in GWs would be a major discovery as it could hardly be interpreted in terms of stellar black holes, allowing for subsequent calibrations of constrained PBH mass functions -- not to mention tests of inflation itself or of any other PBH production mechanism. All this constitutes a strong motivation to consider scenarios in which PBHs would represent a subcomponent of DM and to explore the potential consequences on other candidates, which usually turn out to be highly non-trivial.

The possible co-existence of (self-annihilating) thermal particle DM and PBHs has been considered for quite some time, first as a mean to grow extended halos around PBHs \cite{Ricotti2007,MackEtAl2007a}, then as a potential way to detect or constrain a weakly-interacting massive particle (WIMP) DM component that would accumulate around PBHs as very spiky halos \cite{LackiEtAl2010}. These early studies, however, relied upon DM accretion models based on fluid-like approaches, which therefore mostly addressed radial infall but could not account for angular momentum effects in the building up of DM spikes around PBHs. A breakthrough came more recently with Eroshenko \cite{Eroshenko2016} who directly integrated the Keplerian orbits starting from a non-relativistic Maxwellian distribution of particle DM falling freely onto PBHs, in patches of the universe detached from the Hubble flow. Indeed, in the thermal DM scenario, particles kinetically decouple from the plasma when they are non-relativistic (usually quite after chemical decoupling \cite{HofmannEtAl2001,Bertschinger2006a,BringmannEtAl2007a}), and may therefore fall onto massive point-like objects if they find themselves in their vicinity. Then, the free fall of particle DM onto a PBH starts at kinetic decoupling during radiation domination, hence in a very dense universe. It continues as the total surrounding density decreases and as the gravitational range of the PBH increases accordingly, essentially until matter-radiation equality, leading to the formation of a very compact spike of DM around the PBH. After equality, secondary infall may begin and further accumulate DM over a much larger volume around PBHs, in which case non-linear effects come into play (the mass of accumulated DM becomes of the same order of that of accreting PBHs). Eroshenko's method cannot address this secondary infall period, but the very inner primeval spikes around PBHs are way too dense and too gravitationally bound to be affected by those non-linear effects. Nevertheless, even though delicate to control accurately, the numerical integration of orbits proposed by Eroshenko is very powerful because it incorporates all angular momentum effects. Interestingly, the final spiky DM compact halos around PBHs predicted by this calculation exhibit complex structures with different power-law indices that depend upon the main model parameters: essentially the PBH and WIMP masses, and the kinetic decoupling temperature (the fraction of DM in the form of PBHs further comes into play to set the amplitude of the post-collapse DM density).

Several studies have since then used Eroshenko's formulation to extract constraints on this mixed PBH-particle DM scenario. For instance, ref.~\cite{BoucennaEtAl2018} derived putative constraints on the PBH fraction from the isotropic extragalactic gamma-ray background (EGB), assuming a particle DM component made of WIMPs whose self-annihilation leads to the ``decay'' of spikes into gamma-rays. This analysis was improved in ref.~\cite{CarrEtAl2021c}, which upgraded the determination of scaling relations at large PBH masses. Further discussion on velocity-dependent annihilation was proposed in refs.~\cite{KadotaEtAl2021,ChandaEtAl2022}. Ref.~\cite{AdamekEtAl2019} complemented the picture by providing simulations of spike formation (though only in a tiny corner of parameter space), showing good agreement with the calculation of Eroshenko. Afterwards, the authors of ref.~\cite{GinesEtAl2022}, based on spike profile approximations taken from ref.~\cite{CarrEtAl2021c}, derived constraints from the analysis of CMB data, and found them more stringent than those inferred from the EGB. More recently, some of us \cite{LavalleEtAl2025} have extensively revised the general method to extract constraints from both CMB and EGB data, and provided useful analytical prescriptions to make quick predictions or cross-check numerical calculations.

In the present paper, we re-examine the CMB constraints in detail. Contrary to the previously cited references, we base our predictions upon a revised version of Eroshenko's calculation. Indeed, a fully analytical understanding of the complex structure of spikes was derived for the first time by some of us in ref.\cite{BoudaudEtAl2021} (\paperI~henceforth), where we improved the phase-space integration part of Eroshenko's original development. This correction led to significant changes in the determination of spike profiles in non-negligible parts of the available parameter space, and we anticipate our results to differ from previous work at least from this perspective --- the analytical transcription of this detailed understanding in terms of predictions for annihilation signals in CMB and EGB observations can actually be found in ref.~\cite{LavalleEtAl2025}, already cited above (\paperII\ hereafter).
Besides, we perform a set of dedicated Monte-Carlo Markov Chain (MCMC) analyses of recent data to obtain statistically realistic bounds on this scenario. In particular, we incorporate a full treatment of the energy deposition in the intergalactic medium and its impact on cosmological observables thanks to dedicated state-of-the-art numerical tools.

The paper is organized as follows. In \citesec{sec:recipes}, we recall the bases for the calculation of energy deposition in the medium as a function of redshift, and derive analytical approximations that allow us to clearly interpret and understand our accurate numerical results. We start by reviewing the generic properties of spikes in \citesec{ssec:time}-\ref{ssec:spikes}, then continue by deriving the global annihilation rate for a population of spikes in \citesec{ssec:core}-\ref{ssec:ann_rate}, before writing down the full expression of the energy deposition in \citesec{ssec:deposition}. Our main analysis of CMB data is performed in \citesec{sec:cmb}, where we describe the statistical method used to extract limits both on the PBH fraction assuming fixed WIMP properties (meant to WIMPs supporters), or on the WIMP annihilation cross section assuming fixed PBH properties (to PBHs supporters). These are our main results, which we discuss in detail and compare with other studies also in that section --- we recommend \change{knowledgeable readers} to go straight to it. We conclude and draw perspectives in \citesec{sec:concl}.

\section{From spikes properties to time-dependent energy deposition in the early universe}
\label{sec:recipes}


In this section, we review the main properties of the DM spikes around PBHs, making explicit the parameter dependencies. We start by discussing some timing aspects that were not considered in \paperI, before characterizing the DM profile of the spikes. We conclude this section by determining the time-dependent annihilation rate relevant to derive CMB constraints. We refer the reader to \paperI\ for a detailed explanation of the building up of DM spikes around PBHs, and to \paperII\ for an extensive analytical derivation of the induced annihilation rate -- we will limit ourselves to more phenomenological discussions here.

\subsection{Time ordering}
\label{ssec:time}

DM spikes form around PBHs during the radiation domination epoch after the kinetic decoupling of DM particles, namely as soon as the latter can stream freely throughout the primordial plasma and thereby ``feel'' the presence of BHs. In \paperI, we implicitly assumed that kinetic decoupling takes place after the formation of PBHs, but on general grounds, this needs not be the case. More precisely, this depends on both the PBH collapse time $\tcoll(\mbh)$ (set by its mass $\mbh$) and the DM kinetic decoupling time $\tkd$ (set by the DM-plasma interaction rate). Conventionally, the latter is characterized by the kinetic decoupling temperature $\Tkd$ which, in the thermal DM scenario, is usually lower than the freeze-out temperature. In \paperI, we characterized the kinetic decoupling in terms of a dimensionless parameter $\xkd=\mchi/\Tkd$, for which a rather limited range from $\sim 10^2$ to $\sim 10^4$ is enough to cover a big part of the most natural available parameter space \cite{BringmannEtAl2007a,Bringmann2009}. A straightforward approximate calculation valid in radiation domination gives
\ben
\label{eq:tkd}
\tkd &\simeq & \left\{\frac{4\pi^3}{45}\,\geff(\mchi/\xkd)\right\}^{-1/2}\frac{\xkd^2}{2\,\Mpl}
\left(\frac{\Mpl}{\mchi}\right)^2\\
&\simeq& \frac{2.42\times 10^{-2}\,{\rm s}}{\sqrt{\geff^{\rm kd}}}\left(\frac{\xkd}{10^4}\right)^2
\left(\frac{\mchi}{100\,{\rm GeV}}\right)^{-2}\nn\,,
\een
where $\Mpl=1/\sqrt{\gnewt}$ is the Planck mass, $\gnewt$ the gravitational constant, and $\geff(T)$ the effective number of relativistic degrees of freedom carried by the radiation energy density (see \paperI), with $\geff^{\rm kd}\equiv\geff(\Tkd)$.

Beside, the PBH collapse time can be estimated in various ways, giving results in good qualitative agreement with more rigorous treatments \cite{HaradaEtAl2013}. One can for instance relate the fictitious PBH density over the Schwarzschild radius $R_{\rm S}$ to its mass, and then demand that it correspond to the critical overdensity $\deltac$ susceptible of collapsing to a BH. This gives:
\ben
\rhobh &=& \frac{3}{4\,\pi}\left(\frac{2\,\gnewt}{c^2}\right)^{-3}\mbh^{-2} = (1+\deltac)\left\{ \rho=\frac{3\,H^2}{8\,\pi\,\gnewt}\right\}\nn\\
\Rightarrow &&
\begin{cases}
  \displaystyle \mbh &=  \displaystyle \dfrac{1}{\sqrt{1+\deltac}}\left(\frac{c^3}{\gnewt}\right)\,t
  \simeq \displaystyle 1\,\Msun \, \left(\frac{t}{5.7\times 10^{-6}\,{\rm s}}\right)\\
  \displaystyle\tcoll &=\displaystyle \sqrt{1+\deltac} \left(\frac{\gnewt}{c^3}\right)\,\mbh
  \simeq \displaystyle 5.7\times 10^{-6}\,{\rm s} \,\left( \frac{\mbh}{1\,\Msun}\right)
\end{cases}\,,
\label{eq:tcoll}
\een
where $\gnewt$ is the gravitational constant and $c$ the speed of light. In the numerical evaluation, we have assumed that $\deltac=w_{\rm rad}=1/3$, above which Jeans instabilities prevail \cite{Carr1975}, with the equation of state $w_{\rm rad}$ appropriate to the radiation domination era; we have also used the corresponding expansion rate $H(t)=1/(2\,t)$. Should we instead relate the PBH mass to the mass contained within the instantaneous causal horizon $r_{\rm h}\equiv c/H(t)$, we would find the same equation up to the factor $\sqrt{1+\deltac}$, i.e. a negligible correction.

Since we are interested in the gravitational capture of DM around BHs before matter-radiation equality, we can define the relevant mass range for this study. A quick textbook-like calculation gives the relative scale factor $\aeq$ (relative to its value today) and then the time $\teq$ of matter-radiation equality:
\ben
\label{eq:teq}
\aeq &=& \frac{(1+\frac{7}{8}\Neff(4/11)^{4/3})\,\Omegag^0}{\Omegam^0} = \frac{1}{(1+\zeq)}\simeq 2.9\times 10^{-4}\\
\Rightarrow \teq & \simeq &\frac{4}{3}\left(1-\frac{\sqrt{2}}{2}\right) \frac{\left(\Omegam^0\,\aeq^{-3}  \right)^{-1/2}}{H_0}\simeq 5.1\times 10^4\,{\rm yr}\nn\,,
\een
which allows us to derive the maximal PBH mass potentially relevant to this study, $\mbh^{\rm eq}=\mbh(\teq)\simeq 2.8\times 10^{17}\,\Msun$. In the above equations, $\Omegam^0$ and $\Omegag^0$ represent the relative matter and photon densities today, $H_0$ the current value of the Hubble rate; we have taken $\Neff=3.046$ for the effective number of neutrinos (\eg~\cite{SalasEtAl2016}) and the Planck 2018 values for the cosmological parameters \cite{PlanckCollabEtAl2020a}. The analytical approximation for time around equivalence assumes a simplified two-component cosmology with matter and radiation only --- units of sideral years (yr here) will be used throughout this paper.

In \paperI\ and \paperII, we implicitly assumed that $\tkd>\tcoll$, which we can now translate in the following condition:
\ben
\label{eq:mbhmaxtime}
\mbh < \frac{4.25\times 10^3 \,\Msun}{\sqrt{\geff^{\rm kd}}}\,\left( \frac{\xkd}{10^4}\right)^2\,
\left(\frac{\mchi}{100\,{\rm GeV}}\right)^{-2}\,.
\een
From this equation, we see that some part of the parameter space defined by the triplet $(\mchi,\mbh,\xkd)$, where $\mchi$ goes from $\sim 1$~MeV to $\sim 10$~TeV, and $\mbh$ formally up to $10^{17}\,\Msun$, can actually be such that kinetic decoupling precedes collapse time ($\tkd\leq\tcoll$). This can typically be the case for both heavy DM particles and heavy PBHs. In \paperI, we assumed that $\tkd$ was the initial time of free fall of DM onto PBHs, which is therefore not strictly true in all possible parameter configurations. Fortunately, accounting for a correct time-ordering of events is actually very simple. If $\tff$ denotes the starting point of free fall, then a natural definition is given by:
\ben
\label{eq:tff}
\tff&=&\tff(\mchi,\xkd,\mbh)\equiv{\rm Max}(\tcoll,\tkd)\,,
\een
such that we can replace the original definition of the influence radius of the PBH at kinetic decoupling in \paperI~[Eq.~(2.13)], namely the radius $\rinfl(\tkd)$ inside which an initially comoving and newly free-streaming DM particle could be gravitationally bound to the PBH, with the following substitutions:
\ben
\label{eq:def_rinfl}
\rinfl(t) &=& \left(\etata\, R_{\rm S}\right)^{1/3}\,(c\,t)^{2/3} = \rinfl(a) \simeq \left(\etata\, R_{\rm S}\right)^{1/3}\,(c\,\teq)^{2/3}(a/\aeq)^{4/3}\\
\rinfl(\tkd) &\longrightarrow & \rinfl(\tff)\,;\;\;\;\rtkd\equiv \frac{\rinfl(\tkd)}{R_{\rm S}} \longrightarrow \frac{\rinfl(\tff)}{R_{\rm S}}\nn\,,
\een
where we have first recalled the definition of $\rinfl(t)$, with the numerical factor $\etata=1.086$ (see \paperI). This improved definition ensures $\rtkd\geq 1$ (which was violated for $\tff>\tkd$). Consequently, while we will still use $\tkd$ ($\rtkd$) as a reference time (radius) in the following, one has to bear in mind that $\tff$ supersedes whenever relevant --- in practice, it suffices to consider the post-collapse DM densty at reduced radii $\tilde{r}$ larger than 1.\footnote{Requiring that, at cosmic time $t$, the radius of influence of a BH should exceed its Schwarzschild radius actually translates into an upper bound on $\mbh$ which is a factor $\sqrt{3/\etata}\simeq 1.66$ smaller than the upper limit derived in \citeeq{eq:mbhmaxtime}.} Orders of magnitude are worth recalling:
\ben
\label{eq:rtkd_mag}
\rtkd \simeq 86.87 \left(\dfrac{\geffkd}{10}\right)^{-1/3} 
\left\{  \left(\dfrac{\xkd/10^4}{\mchi/100\,{\rm GeV}}\right)=\left(\dfrac{\Tkd}{10\,{\rm MeV}}\right)^{-1}\right\}^{4/3}\left(\dfrac{\mbh}{1\,\Msun}\right)^{-2/3}\,,
\een
where $\rtkd\leq 1$ would roughly indicate that the BH collapses after kinetic decoupling, which also means that the DM density at actual collapse time, the one relevant to set the DM spike density, would accordingly be smaller than the overall DM density at kinetic decoupling. Finally, we define the maximal radial extension of the spike by the influence radius at matter-radiation equivalence:
\ben
\label{eq:rtspike}
\rtspike = \rteq = \rtinfl(\teq) = 2.89\times 10^{11}\,
\left(\dfrac{\teq}{5.15\times 10^4\,{\rm yr}}\right)^{2/3}
\left(\dfrac{\mbh}{1\,\Msun}\right)^{-2/3}\,.
\een

\subsection{Summary of spikes properties}
\label{ssec:spikes}
The steep density profiles (dubbed spikes) of particle DM that develop around PBHs in the radiation era were shown by Eroshenko \cite{Eroshenko2016} to follow radial power laws, pointing to some scale invariance fixed by the main physical parameters, $\mbh$, $\mchi$ and $\Tkd$. In \paperI, we fully determined the origin of these power laws from analytical calculations. To summarize, except for a very peculiar regime for which caustics give rise to an index of $\gamma=3/4$ in the very inner parts of spikes (typically when $\tilde{r}\lesssim \xkd$ and for light BHs), one can mostly distinguish two main logarithmic slopes: (i) $\gamma=3/2$ which is essentially universal,\footnote{Note that as explained in \paperI, the 3/2 index is found in two completely different situations: (i) when the surrounding DM is thermal-pressure dominated (for relatively light BHs and DM particle escape speed smaller than dispersion velocity), and (ii) for DM particles strongly bounded to the BH inside the influence radius at kinetic decoupling (for heavier BHs). Our phenomenological discussion will mostly be based on the former case -- for more details, we refer the reader to \paperII.} transitioning to (ii) $\gamma=9/4$ at larger radii (earlier on as the BH mass increases). This transition mostly delineates two different regimes, which will turn out to be the most relevant in this study: a kinetic pressure dominated system, and a gravitational potential dominated system at the onset of DM free fall. Recalling that the gravitational influence radius of the BH can be traded for time or scale factor (see \citeeqp{eq:def_rinfl}), a transition radius $\rbreak$ can then be defined such that:
\ben
\dfrac{G\,\mbh}{\rbreak} = \dfrac{\langle v^2\rangle(\rbreak)}{2}\,,
\een
up to a factor of 2 depending on whether one takes null energy condition or virial theorem as a proxy. Here, the (squared) particle velocity dispersion $\langle v^2\rangle $ relates to the kinetic decoupling temperature and the scale factor, hence to a distance of $\rbreak$ to the BH through the relation:
\ben
\langle v^2\rangle = \langle v^2\rangle(a_\mathrm{b}) = \dfrac{3}{\xkd}\left(\dfrac{a_\mathrm{b}}{a_\mathrm{kd}}\right)^{-2} \simeq \dfrac{3}{\xkd}\left(\dfrac{\rbreak}{\rkd}\right)^{-3/2} = \langle v^2\rangle(\rbreak)\,,
\een
whose evolution in terms of scale factor is made explicit, and where we have traded the scale factor for the influence radius by means of \citeeq{eq:def_rinfl} --- for simplicity, we have neglected the mild dependence in relativistic degrees of freedom in the previous step. We can therefore extract an expression\footnote{In \paperI, a similar expression, $\rtbreak = \xi^2\sigma_{\rm kd}^4\rtkd^3$, was found, with $\sigma_{\rm kd}^2=1/\xkd$ and $\xi$ a parameter taking values between $\sim 2.4$ and 3.2.} for the dimensionless form of the transition radius:
\ben
\label{eq:rbreak}
\rtbreak = 9\,\xkd^{-2} \,\rtkd^3\, 
\approx 5.9\times 10^{-2} 
\left( \dfrac{\geffkd}{10} \right)^{-1}
\left( \dfrac{\xkd}{10^4} \right)^2
\left( \dfrac{\mchi}{100\,{\rm GeV}} \right)^{-4}
\left( \dfrac{\mbh}{1\,\Msun} \right)^{-2}
\,,
\een
where the final scaling in terms of $\xkd$, $\mchi$ and $\mbh$ derives from \citeeq{eq:rtkd_mag}. It notably explains why for light WIMPs and/or light BHs, the 3/2 slope can extend up to whole spike radius, in which case the 9/4 slope never develops (full kinetic pressure domination). The approximate condition to witness a transition from the 3/2 to the 9/4 slope then translates into the condition $\rtbreak\lesssim \rtspike=\rteq$, which gives:
\ben
\left(\dfrac{\geffkd}{10}\right)^{-1}
\left(\dfrac{\xkd}{10^4}\right)^2 
\left(\dfrac{\mchi}{100\,{\rm GeV}}\right)^{-4} 
\left(\dfrac{\mbh}{3.0\times 10^{-10}\,\Msun}\right)^{-4/3} 
\left(\dfrac{\teq}{5.15\times 10^4\,{\rm yr}}\right)^{-2/3}
\lesssim  1\,.\nn\\
\label{eq:break_condition}
\een
This condition is essentially fulfilled for both heavy BHs and heavy WIMPs, while it is not for BHs lighter than $\sim 10^{-11}\,\Msun$, for which the 3/2 slope can therefore extend up to $\rtspike$. For very light WIMPs with $\mchi\ll 1\,{\rm GeV}$, the 9/4 slope even never shows up all over the relevant BH mass range. This corresponds to a spike formation regime fully dominated by kinetic pressure.

On the other hand, the slope of 3/2 (and consequently 3/4) may never develop in the opposite case, \ie~when the dynamics is fully dominated by the BH gravitational potential (typically for heavy BHs) -- the profile is then entirely described by the 9/4 slope. This configuration corresponds to yet another approximate condition, $\rtbreak\lesssim 1$, which translates into:
\ben
\left(\dfrac{\geff^{\rm kd}}{10}\right)^{-1}
\left(\dfrac{\xkd}{10^4}\right)^{2}
\left(\dfrac{\mchi}{100\,{\rm GeV}}\right)^{-4}
\left(\dfrac{\mbh}{2.4\times 10^{-1}\,\Msun}\right)^{-2}
\lesssim 1
\een
Therefore, except for light relic particles, the slope of 9/4 characterizes most of the spikes for BHs heavier than a few $\Msun$.

We can now try to turn this short summary into an empirical modeling, in a way slightly improved with respect to refs.~\cite{CarrEtAl2021c,GinesEtAl2022}, and meant to capture all of the main parameters dependencies. The onset of DM free fall extends from $\tff$ to $\teq$, which corresponds to the gravitational influence radius growing from $\rkd$ to $\req$, respectively, during which a multi-slope spike develops. If $\xkd\lesssim \rtkd$, an inner 3/4 logarithmic slope can form at radii $\tilde{r}\leq \xkd$, followed by a 3/2 slope outside. A further transition from 3/2 to 9/4 can occur whenever $\xkd<\rtbreak<\rteq$. This can be formulated in terms of a very approximated picture of the spike density distribution as a function of the dimensionless radius $\tilde{r}$, which, from external toward inner parts, reads:
\ben
\label{eq:rhospike_approx}
\rhospike(\tilde{r}) \approx
\begin{cases}
(1-\fbh)\rhocdm^\mathrm{eq} \,\left(\dfrac{\tilde{r}}{\rteq}\right)^{-9/4} \;\; & \text{if}\;\rtbreak\leq \tilde{r}< \rteq \\
(1-\fbh)\rhocdm^\mathrm{eq} \,\left(\dfrac{\rtbreak}{\rteq}\right)^{-9/4}\,\left(\dfrac{\tilde{r}}{\rtbreak}\right)^{-3/2} & \text{if}\;\xkd\leq \tilde{r}< \rtbreak \\
(1-\fbh)\rhocdm^\mathrm{eq} \,\left(\dfrac{\rtbreak}{\rteq}\right)^{-9/4}\,\left(\dfrac{\xkd}{\rtbreak}\right)^{-3/2} \,\left(\dfrac{\tilde{r}}{\xkd}\right)^{-3/4} & \text{if}\; 1 \leq \tilde{r}< \xkd  < \rtkd
\end{cases}
\een
where $\rhocdm^\mathrm{eq}$ is the total DM density at matter-radiation equality, $\fbh$ the fraction of DM in BHs. Here we the peculiar 3/4-slope regime, which can show up if $\xkd \lesssim \rtkd$, is made explicit for completeness even though it will actually never play any role in the following calculations.

This very crude expression, whose normalization approximates the true one up to a factor of order unity (see \paperII), allows us to recover the main features of the spikes, in particular the fact that the normalization densities for the slopes 3/4 and 3/2 are mostly independent of the BH mass, and therefore universal.\footnote{One of the two 3/2 regimes does actually depend on the BH mass -- see \paperII.} Indeed, in spite of the rather sibylline combination of scale radii, this can be seen by noticing that the product $(\rtbreak/\rteq)^{-9/4}\,\rtbreak^{3/2}$ is independent of the BH mass. The above expression also tells us that for BHs heavy enough so that $\rtbreak \lesssim 1$, for given sets of the other key parameters [see \citeeq{eq:rbreak}], the only relevant slope is 9/4.

We emphasize that this empirical expression will just be used for a qualitative understanding. All of the following illustrations and results will rely upon accurate numerical calculations of density profiles, unless specified otherwise.

\subsection{Time-dependent annihilation core}
\label{ssec:core}
Since we are considering self-annihilating DM, the huge densities accumulated in spikes cannot remain unaffected by self-annihilation, and the latter actually regulates the amount of DM right around the BH. Density cores arise from this process, which we try to characterize below.

A crude way to describe this process is to consider a simple evolution equation assuming circular orbits \cite{AhnEtAl2007,Vasiliev2007}:
\ben
\dfrac{\dd n_\chi(r,t)}{\dd t} = -2\times\dfrac{\deltaann \sigvav}{2}\,n_\chi^2(r,t) = 
- \dfrac{n_\chi(r,t)}{\tann(n_\chi)}\,,
\een
where $\deltaann=1$ (1/2) for Majorana (Dirac) particles, and the first factor of 2 accounts for the disappearance of a pair of particles in an annihilation process. The characteristic WIMP annihilation time is given by:
\ben
\label{eq:tann}
\tann \equiv \left\{ \left( n_\chi= \frac{\rho_\chi}{\mchi}\right) \sigvav \right\}^{-1}\,,
\een
where $\mchi$ is the WIMP mass, $n_\chi$ its number density, $\rho_\chi$ its corresponding mass density, and $\sigvav$ the averaged cross-section times relative velocity. From now on, we will concentrate on Majorana species, hence $\deltaann =1$. The above differential equation finds a trivial solution that we can express in terms of the mass density:
\ben
\label{eq:rhomax_exact}
\rhomax(r,\Delta t) = \dfrac{\rhomax^0(\Delta t)}{\left(1+\rhomax^0(\Delta t)/\rho_\chi^0(r)\right)} 
\overset{\rho_\chi^0\gg\rhomax^0}{\looongrightarrow} 
\rhomax^0(\Delta t) \equiv \dfrac{\mchi}{\sigvav \,\Delta t }\,, 
\een
where $\rho_\chi^0=\rho_\chi(r,t_0)$ represents the initial value of the density profile at time $t_0$, and $\Delta t$ the time elapsed since then. In the following, we will take $t_0=\teq$ (taking $\tff$ instead would not impact the results which consider the spikes densities at much later times). We emphasize that this crude treatment relies either on the fully circular or on the fully radial orbit assumption. Extensions of this approximation to more general orbits and to more general phase-space distribution functions were discussed in \eg~refs.~\cite{Vasiliev2007,ShapiroEtAl2016}, but we leave a more refined revision of this to future work. We note that such extensions generically lead to higher values for this saturation density, which makes the current approximation rather conservative. We also note that only using the asymptotic limit $\rhomax^0$ in \citeeq{eq:rhomax_exact} can lead to an error of up to a factor of $\sim 2$ in the spike annihilation rate with respect to the exact expression featuring the initial density at time $t_0$. We will therefore use the later in our numerical calculations.

To summarize, if too large, the WIMP density around a PBH is skimmed until $\tann$ becomes of the order of the age of the spike $\Delta t$. An order of magnitude estimate is given below:
\ben
\label{eq:rhomax}
\rhomax(\Delta t)\simeq
\begin{cases}
  4.57\times 10^{-10}\,{\rm g/cm^3}\\
  2.57\times 10^{14}\,{\rm GeV/cm^3}
\end{cases}
\left(\frac{\mchi}{100\,{\rm GeV}}\right)
\left(\frac{\sigvav}{3\times10^{-26}{\rm cm^3/s}}\right)^{-1}
\left(\frac{\Delta t/\teq}{8}\right)^{-1}\,.\nn
\een
This maximal density, which here is calculated at a time $\Delta t/\teq = 8$ (close to recombination), is still $\sim 10^{10}$ times as large as the total DM density at matter-radiation equality, and characterizes an almost homogeneous core around the PBH with a size that depends on the spike profile.

In the following, without loss of generality, we describe the initial spike profile crudely approximated in \citeeq{eq:rhospike_approx} as a generic power-law in radius of index $\gamma$:
\ben
\label{eq:rhospike_simple}
\rho_{\rm spike}^{\rm ini}(\tilde{r}) =(1-\fbh)\,\rhogamma \left(\frac{\tilde{r}}{\rtgamma}\right)^{-\gamma}\,,
\een
where $\rhogamma$ and $\rtgamma$ can be related to the normalization factors introduced in \citeeq{eq:rhospike_approx}, and where we keep track of the BH fraction $\fbh$. Due to self-annihilation, this density profile may therefore saturate as time goes on if it exceeds values  $\sim \rhomax(t)$, leading to the formation of a density core of size $\rcore$ that grows with time as:
\ben
\label{eq:rcore}
\rtcore(t) &\equiv & \rtgamma \left\{\frac{(1-\fbh)\,\rhogamma}{\rhomax(t)}\right\}^{\frac{1}{\gamma}}
\propto \Delta t^{\frac{1}{\gamma}}\,.
\een
In this description, the growth of $\rcore$ with time would stop if it exceeded the size of the spike (here we conservatively neglect the more extended halo that is expected to develop beyond the spike from secondary infall). Summing up from the definitions of the core radius $\rcore$ and of the core density $\rhomax$ introduced above, and assuming the core remains smaller than the spike, the actual profile of the spike at the direct vicinity of the PBH can be reformulated and decomposed into two parts:
\ben
\label{eq:spike}
\rho_{\rm spike}(\tilde{r}) =
\begin{cases}
    \rhomax(t)\;\;\; & \forall \; \etaS  < \tilde{r} \leq \rtcore\\
  \rhomax(t) \left(\dfrac{\tilde{r}}{\rtcore(t)}\right)^{-\gamma}& {\rm otherwise}
\end{cases}\,.
\een
We have introduced parameter $\etaS$, which allows us to define the distance to the BH (in Schwarzschild unit) below which Keplerian motion could be subject to relativistic corrections (all orbits with pericenters getting close to $R_{\rm S}$) -- we are not referring to relativistic effects affecting spikes forming from the adiabatic growth of central BHs, as the formation process is completely different here. This is a free parameter, of order 1-5 \cite{SadeghianEtAl2013}, below which we very conservatively set the DM density to 0 (though it is actually expected to increase further). This parameter may be used to quantify theoretical uncertainties -- unless specified otherwise, it is set to 1 by default (we checked our results do not change if we take $\etaS=3$ instead).

An important aspect of the above profile is its time dependence, owing to the fact that the maximal density decreases and the core radius increases non trivially with time. This will help us carefully determine the WIMP annihilation rate around PBHs in the next paragraph. It is also interesting to note that the only dependence on $\fbh$ is hidden in the definition of $\rtcore(t)$.
\begin{figure}[t!]
\centering
\includegraphics[width=\linewidth]{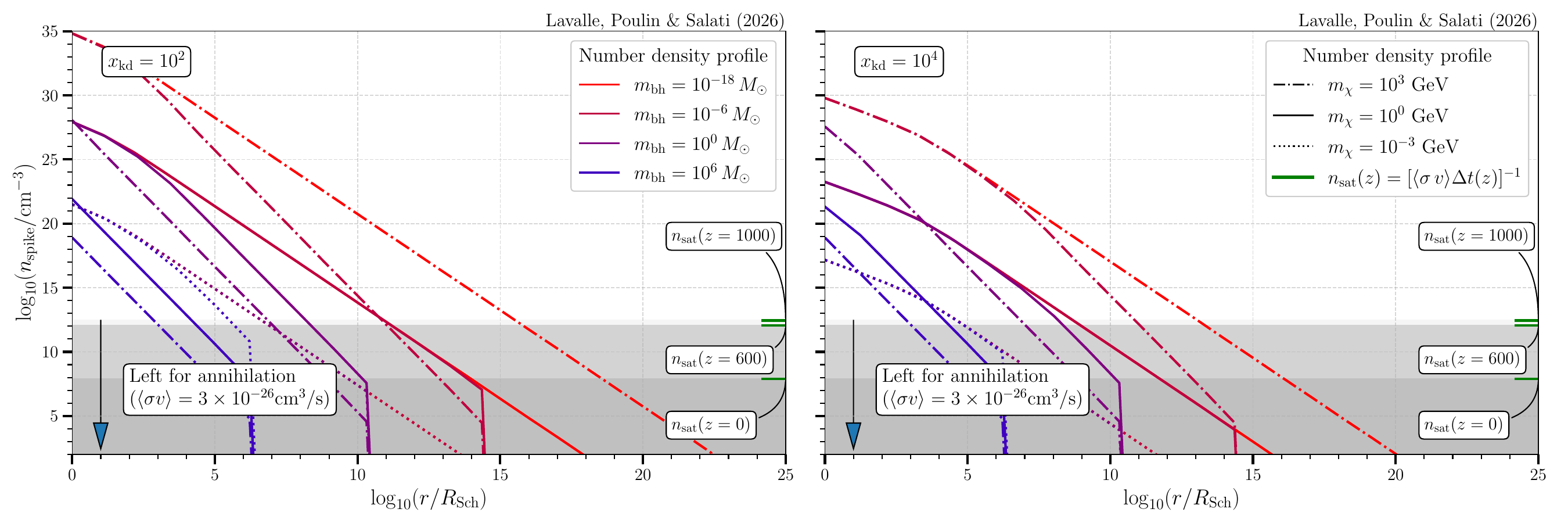} 
\caption{Spikes number density profiles for different combinations of parameters, \ie~$\mbh\in[10^{-18}-10^6]\,\Msun$ (colors), and $\mchi\in[10^{-3}-10^3]\,{\rm GeV}$ (line styles). What remains out of self-annihilation is located in the grey bands (from $z=1000$ to 0 downward). {\bf Left panel:} early kinetic decoupling ($\xkd=10^2$). {\bf Right panel:} late kinetic decoupling ($\xkd=10^4$).}
\label{fig:sat_density}
\end{figure}

We illustrate the level of the saturation density in its asymptotic (and therefore approximate) expression against different spike configurations (varying $\mchi$ and $\mbh$), and at different redshifts, in \citefig{fig:sat_density}. We see (i) the large diversity of profiles stemming from the different input parameters, and (ii) that already at redshift $z=1000$, large fractions of spikes have annihilated away (regions still available for self-annihilation are located in the gray bands -- 3 bands corresponding from top to bottom to $z=1000$, 600, and 0). The left (right) panel assumes an early (late) kinetic decoupling with parameter $\xkd=10^2$ ($10^4$). Note that we show the number density profiles rather than the mass density profiles to be able to define a single saturation density per redshift (it then only depends on time and cross section, which we fix here to $\sigvav=3\times 10^{-26}\,{\rm cm^3/s}$). From this figure, it is rather clear that the very inner parts of profiles exhibiting a 3/4 behavior (typically for $\mbh<1\,\Msun$ in the plots) are irrelevant because annihilated away. It is also clear that parts of the profiles left beyond the saturation core will either decrease like $\tilde{r}^{-3/2}$ (light BHs) or $\tilde{r}^{-9/4}$ (heavy BHs), with a smooth transition between these two regimes.

\subsection{Time-dependent WIMP annihilation rate for a population of spikes (or global spike decay rate)}
\label{ssec:ann_rate}

A crucial step is to establish the overall WIMP annihilation rate, which is expected to be strongly boosted in the presence of early``accreting" PBHs \cite{MackEtAl2007a,LackiEtAl2010,Eroshenko2016,BoucennaEtAl2018,AdamekEtAl2019,GinesEtAl2022}. Improving over previous studies, we will consider carefully the time dependence of that annihilation rate, as we will deal with observational constraints arising at very different epochs of the universe.

\subsubsection{Annihilation rate around a single PBH (or single spike decay rate)}
\label{ssec:gamma_bh}
We start by considering the annihilation rate around a single BH. It may be expressed as an effective decay rate for the BH, even though it is only the spike that decays:
\ben
\label{eq:gammabh}
\Gammabh(t) &=& 
\frac{\sigvav}{2}R_{\rm S}^3
\int_{\etaS }^{\rtspike} \dd^3\vec{\tilde{r}}\left(\frac{\rho_{\rm spike}(\tilde{r},t)}{\mchi}\right)^2\\
&\simeq& 
\frac{\sigvav}{2}
\frac{4\pi}{3} R_{\rm S}^3\,(\rtcore(t))^3\left(\frac{\rhomax(t)}{\mchi}\right)^2\nn\\
&& \times \left\{
 \left(1-(\rtcore/\etaS)^{-3}\right)
+ \frac{3}{2\gamma-3}\left(1-(\rtspike/\rtcore)^{3-2\gamma}\right)
\right\}\,,\nn
\een
where, in the second line, we have taken the approximate cored power-law profile of index $\gamma$ given in \citeeq{eq:spike}, and where $\tilde r_{(\ldots)}\equiv r_{(\ldots)}/R_{\rm S}$ is the dimensionless version of radii in units of Schwarzschild radius. For the moment, we implicitly assume that $\rtcore\leq\rtspike$, which is almost always the case. In the last expression in brackets, the left-hand-side term characterizes the annihilation yield from the core, while the right-hand-side term accounts for the contribution of the external part outside the core. It turns out that $\rtcore\gg \etaS \sim 1$ to a good approximation, so the former term barely deviates from $\sim 1$. The strength of the latter term depends on  index $\gamma$, which mostly takes a value of 3/2 or 9/4, as discussed above. Hence, we can approximate the effective decay rate as follows:
\ben
\label{eq:gammabh_eff}
\Gammabh(t) &\simeq & 
\frac{2\pi\,K_\gamma}{3}\, \sigvav \left(\frac{\rhomax(t)}{\mchi}\right)^2 R_{\rm S}^3 \,\left(\rtcore(t)\right)^3
\propto  \Delta t^{3/\gamma-2}
\,,
\een
where, for the moment, we only put forward the time dependence originating from the expressions of $\rhomax$ and $\rtcore$. There is also a tiny residual time dependence in the pseudo-constant $K_\gamma$ introduced just above, that we more formally define as:
\ben
\label{eq:kgamma}
K_\gamma \equiv
\begin{cases}
  1-(\rtcore/\etaS)^{-3}+3\,\ln(\rtspike/\rtcore(t)) & \text{for $\gamma=3/2$ and $\rtcore(t)<\rtspike$}\\
  1-(\rtcore/\etaS)^{-3}+2\left(1-(\rtspike/\rtcore(t))^{-3/2}\right) \;\;\; & \text{for $\gamma=9/4$ and $\rtcore(t)<\rtspike$}\\
  \left((\rtspike-\etaS)/\rtcore(t)\right)^3 & \text{if $\rtcore(t)\geq\rtspike$}
\end{cases}
\een
This pseudo-constant $K_\gamma$ is of order $\sim 1$-10. Note that although this only concerns a vanishingly small portion of the available parameter space, we have explicitly considered the case in which $\rtcore(t)$ exceeds the radial extend of the spike, $\rtspike$. In this case, the density core stops expanding with time even though the maximal density keeps on decreasing, leading to $\Gammabh(t)\propto \sigvav \rhomax^2 \rtspike^3 \propto \sigvav^{-1}\Delta t^{-2}$ instead of $\propto (\sigvav \Delta t)^{3/\gamma-1}/\Delta t$ --- this is the manifestation of the incompleteness of our spike modeling, which neglects the DM accretion beyond the spike radius that should follow from secondary DM infall after matter-radiation equality. This can still be considered as a very conservative assumption when it comes to observational constraints, because all contributions to the annihilation yield beyond the spike radius fixed at equivalence are neglected.


\begin{figure}[t!]
\centering
\includegraphics[width=\linewidth]{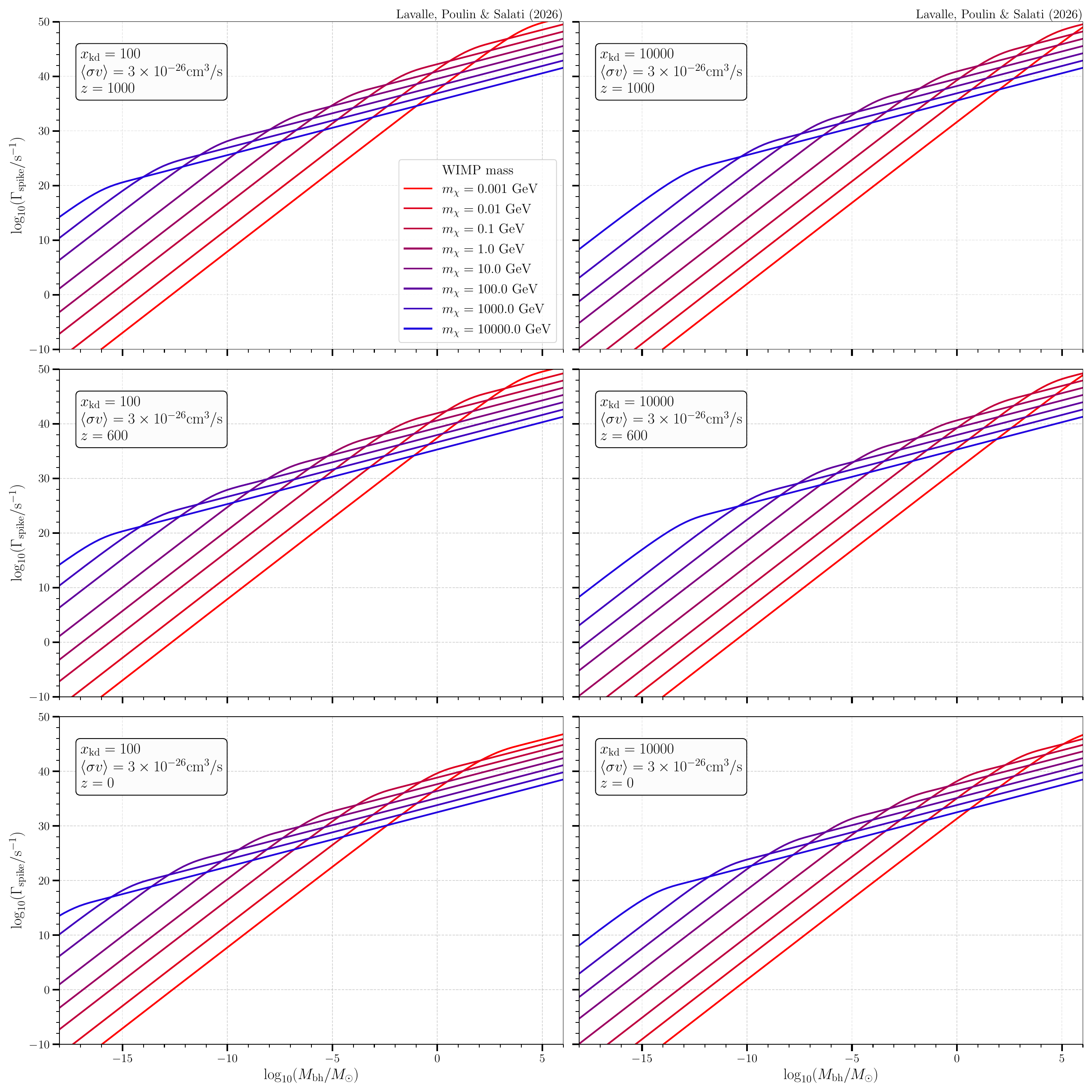} 
\caption{Single spike decay rate (or annihilation rate in single spikes) as a function of BH mass, for different WIMP masses (from 1~MeV to 10~TeV -- line color code), and at different redshifts: $z=1000$ (top panels), $z=600$ (middle panels), and $z=0$ (bottom panels) -- a BH fraction of $\fbh=10^{-3}$ is assumed, as well as an annihilation cross section of $\sigvav=3\times 10^{-26}\,{\rm cm^3/s}$.  {\bf Left column:} early kinetic decoupling ($\xkd=10^2$). {\bf Right column:} late kinetic decoupling ($\xkd=10^4$).}
\label{fig:gamma_bh_single}
\end{figure}

We show the results of the full numerical calculation of single spikes decay rates (\ie~WIMP annihilation rate in single spikes) as functions of BH mass in \citefig{fig:gamma_bh_single}, for different WIMP masses (from 1~MeV to 10~TeV, red to blue curves), at different redshifts ($z=1000$ to 0 from top to bottom panels), and for early (late) kinetic decoupling in left (right) panels.

The first important thing to notice is that predictions span over $\sim 60$ orders of magnitude as the BH mass goes from $10^{-18}$ to $10^6\,\Msun$, for light WIMPs.
The second important feature of \citefig{fig:gamma_bh_single} is the scaling of $\Gammabh$ which goes $\propto \mbh^3$ at low BH mass, and transitions to $\propto \mbh$ at some specific breaking BH mass -- this was already noticed in ref.~\cite{BoucennaEtAl2018}. This scaling relation is actually related to the transition from the slope of 3/2 to that of 9/4, as discussed around \citeeq{eq:rhospike_approx}.
We can understand it by inserting the full expression of the core radius $\rtcore(t)$ given in \citeeq{eq:rcore} into the simplified expression of the spike decay rate given in \citeeq{eq:gammabh_eff}, which reveals the dependence on the relevant scale density $\rhogamma$ and radius $\rtgamma$ of our generic spike profile of index $\gamma$ [see \citeeq{eq:rhospike_simple}]:
\ben
\label{eq:gammabh_eff_expl}
\Gammabh(t) &\propto & 
\rhomax^2(t)\, R_{\rm S}^3 \, (\rtcore)^3
\propto \rhomax^2(t)\, R_{\rm S}^3 \,
(\rtgamma)^3 \left\{\frac{(1-\fbh)\,\rhogamma}{\rhomax(t)}\right\}^{\dfrac{3}{\gamma}}
\,.
\een
For completeness, let us write down explicitly the expressions of the normalization density $\rhogamma$ and of the scale radius $\rtgamma$ as inferred from \citeeq{eq:rhospike_approx} for the relevant indices $\gamma$ (3/2 and 9/4):
\ben
\rhogamma = 
\begin{cases}
 \rho^0_{3/2} \approx \rhocdmeq \left(\frac{\rtbreak}{\rteq}\right)^{-9/4} \\
\rho^0_{9/4} \approx \rhocdmeq 
\end{cases}
\;\text{and}\;
\rtgamma = 
\begin{cases}
\tilde{r}^0_{3/2}\approx \rtbreak\\
\tilde{r}^0_{9/4}\approx \rteq
\end{cases}
\,.
\een
With these expressions in mind and inspecting further \citeeq{eq:gammabh_eff_expl}, we note that the saturation density $\rhomax$ is essentially independent of the BH mass. Besides, as discussed below \citeeq{eq:rhospike_approx}, for $\gamma=3/2$, a slope characteristic of the light to moderate BH mass regime, the normalization density $\rhogamma \times \rtgamma{}^\gamma$ does not depend on the BH mass, hence neither $\rtcore(t)$. Consequently, the approximate analytical expression of the spike decay rate in \citeeq{eq:gammabh_eff} becomes
$\Gammabh\propto\rhomax^2\, R_{\rm S}^3 \,\rtcore{}^3 \propto\mbh^3$. As the BH mass increases, the break radius $\rtbreak\propto\mchi^{-4}\,\mbh^{-2}$ decreases (see \citeeqp{eq:rbreak}) and becomes smaller than the core radius. This corresponds to the transition of the spike profile from a 3/2 to a 9/4 slope, at which, still consistently with \citeeq{eq:rhospike_approx}, the scale radius becomes $\rtgamma = \rteq \propto \mbh^{-2/3} $, while the scale density remains independent of the BH mass. Therefore, the core radius inherits the very same dependence owing to the above equation, such that in this regime $\Gammabh\propto\rhomax^2\, R_{\rm S}^3 \,\rtcore{}^3 \propto \mbh$. The transition between these two regimes occurs when $\rtcore\sim\rtbreak$. Using the 9/4 profile, this translates into a breaking BH mass:
\ben
\mbh^{\rm b} &\approx& 5.3\times 10^{-7}\,\Msun\,
(1-\fbh)^{-1/3}
\left(\dfrac{\xkd}{10^4}\right)^{3/2}
\left(\dfrac{\mchi}{100\,{\rm GeV}}\right)^{-8/3}
\left(\dfrac{\sigvav}{3\times 10^{-26}\,{\rm cm^3/s}}\right)^{-1/3}\nn\\
&\times &
\left(\dfrac{\geffkd}{10}\right)^{-3/4}
\left(\dfrac{\rhocdm(z_{\rm eq})}{8.82\times 10^{-20}\,{\rm g/cm^3}}\right)^{-1/3}
\left(\dfrac{\teq}{5.15\times 10^4\,{\rm yr}}\right)^{-1/2}
\left(\dfrac{\Delta t/\teq}{8}\right)^{-1/3}
\nn\\
\label{eq:mbreak}
\een
which is evaluated at a time $\Delta t/\teq = 8$ close to recombination. When compared with \citefig{fig:gamma_bh_single}, it is rather clear that this expression captures reasonably well the actual position of the break (see \paperII\ for more technical details).

We are armed enough to estimate the scaling relations obeyed by the spike decay rate in different regimes, which simply amounts to plugging the expression of the core radius of \citeeq{eq:rcore} in \citeeq{eq:gammabh_eff}. In the 3/2 case (\ie~very light BHs, $\mbh\ll\mbh^{\rm b}$), we find:
\ben
\label{eq:gammabh_32}
\Gammabh^{(3/2)} &\approx& \frac{2\pi}{3}\,
K_{3/2} \, \sigvav \,
\left\{ \frac{(1-\fbh)\,\rhocdmeq }{\mchi} \right\}^2 \, \RSch^3 \, \left\{ \rtbreak = 9\,\xkd^{-2}\,\rtkd^3 \right\}^{-3/2}\,\rteq^{9/2} \\
&\approx&
\dfrac{2\pi}{81}\,K_{3/2} \, (1-\fbh)^2\, \sigvav \,
\left(\dfrac{\rhocdmeq}{\mchi}\right)^2\,
\xkd^{3}\, \RSch^3\, \left( \dfrac{\teq}{\tkd}\right)^3 \nn\\
&\approx & 1.61 \times 10^{-2}\,{\rm s^{-1}} \,\times 
\left[\frac{K_{3/2}}{5}\right]\,  (1-\fbh)^2\,
\left[\frac{\mbh}{10^{-18}\,\Msun}\right]^3\nn\\
&&\times \left[\frac{\sigvav}{3\times 10^{-26}\,{\rm cm^3/s}}\right] \,
\left[\frac{\geffkd}{100}\right]^{3/2}\,
\left[\frac{\xkd}{10^4}\right]^{-3}\,
\left[\frac{\mchi}{100\,{\rm GeV}}\right]^4\,.
\nn
\een
The accurate calculation, accounting for the precise number of relativistic degrees of freedom, gives $\sim 6\times 10^{-2}$~s$^{-1}$ (roughly independent of redshift), which confirms that the order-of-magnitude estimate is correct. From this relation, on top of the BH mass dependence $\propto \mbh^3$, we also obtain two additional important scaling relations, (i) $\propto \mchi^4$, which matches perfectly with the relative ratios observed between the curves in the left part of all panels in  \citefig{fig:gamma_bh_single}, and (ii) $\propto \xkd^{-3}$, which explains the relative amplitudes between the curves of left panels and those of right panels (change in $\xkd$). We also emphasize that in this regime, the decay rates barely depend on time (a slight logarithmic dependence is hidden in $K_\gamma$, but almost irrelevant here). This is due to the fact that in the 3/2-slope case, the decrease of the squared annihilation-induced core density is exactly compensated by the increase of the core volume. On the other hand, the interpretation of the scaling $\propto \xkd^{-3}$ is a bit less straightforward, even though it makes sense that earlier kinetic decoupling, which occurs in a denser universe, should lead to denser spikes. Actually, that scaling comes from the fact in the Keplerian 3/2 regime (pressure dominated dynamics), the density at the vicinity of the BH within a radius $\rtkd$ reflects the invariant phase-space density set by Liouville's theorem, and scales like $\propto \rho_{\rm dm}^{\rm kd}/\sigma_{\rm kd}^3\propto \xkd^{-3/2}$ --- taking the square for the annihilation rate leads to the correct scaling in $\xkd$ (see \paperI\ and \paperII\ for more technical details).

In the 9/4 case (\ie~heavy BHs with $\mbh>\mbh^{\rm b}$), new scaling relations are obtained, which read:
\ben
\Gammabh^{(9/4)} &\approx& 
\frac{8\,\etata}{9} \left(1-\dfrac{\sqrt{2}}{2}\right)^2\,  K_{9/4}\,
 \left(\frac{(\rhocdmeq)^{4/3}}{\rho_{\rm m}^{\rm eq}}\right) \,
 (1-\fbh)^{4/3}\,\mbh\,\mchi^{-4/3}\,\sigvav^{1/3}\, \Delta t^{-2/3}
\nn\\
&\approx& 2.04\times 10^{44}\, {\rm s^{-1}}\times 
\left[\frac{K_{9/4}}{5}\right]\,
\left[\frac{\Delta t/\teq}{8}\right]^{-2/3}\,
\left[\frac{\teq}{5.15\times 10^{4}\,{\rm yr}}\right]^{-2/3}\nn\\
&&\times
(1-\fbh)^{4/3}\,
\left[\frac{\mbh}{10^6\,\Msun}\right]\,
\left[\frac{\mchi}{100\,{\rm GeV}}\right]^{-4/3}\,
\left[\frac{\sigvav}{3\times 10^{-26}\,{\rm cm^3/s}}\right]^{1/3}\,,
\label{eq:gammabh_94}
\een
to be compared with the accurate numerical prediction of $1.69\times 10^{44}$~s$^{-1}$ at redshift 1100 ($8.62\times 10^{43}$~s$^{-1}$ at redshift 600), again reasonably close. Here, on top of the transition to a dependence $\propto \mbh$ already emphasized, we notably see a reversal in the WIMP mass dependence beyond the 9/4 break with respect to the 3/2 case. Now, indeed, lighter WIMPs contribute higher spike annihilation rates. Another striking change with respect to the 3/2 slope regime is that the kinetic decoupling pseudo-temperature $\xkd$ stops playing any role. This is consistent with the fact that in this regime, the spike dynamics is entirely set by the BH's gravitational potential such that initial kinetic pressure has no influence. Finally, a time dependence $\propto \Delta t^{-2/3}$ shows up owing to that the squared core density decreases faster than the core volume in the 9/4 regime.

All the discussion above mostly assumes that the annihilation core radius $\rtcore(t)$ is always smaller than the spike radius $\rtspike=\rteq$. However, as already mentioned below \citeeq{eq:kgamma}, this needs not be the case, and there is actually a value of the product $\sigvav \Delta t$ above which that assumption breaks down. It can precisely be found by equating $\rtcore(t)$ and $\rtspike$:
\ben
\left\{ \sigvav \,\Delta t\right\}_{\rm max} &=&
\left( \dfrac{\rtgamma}{\rtspike} \right)^{-\gamma}\,
\dfrac{\mchi}{(1-\fbh)\,\rhogamma}\\
&\approx& 6.49\times 10^{-20} \,{\rm cm^3/s\cdot Gyr}\,
(1-\fbh)^{-1}\left( \dfrac{\mchi}{100\,{\rm GeV}}\right)\nn\,,
\een
where the numerical evaluation is here performed in the 9/4 regime for illustration. Above this value, the signal starts decreasing linearly with increasing product $\sigvav \Delta t$.


\subsubsection{Annihilation rate around a population of PBHs}
\label{sssec:ann_pop}
Now we turn to the evaluation of the collective annihilation rate induced by a population of PBHs. In the following, we assume a monochromatic mass function for PBHs, but a generalization to an extended mass function would be straightforward. We also assume that PBHs make up a fraction $\fbh$ of DM, so the annihilation rate per unit volume can be written as:
\ben
\label{eq:gammabhtot}
\frac{\dd \Gammabht(\mbh,\mchi,\tff,t)}{\dd V} = \left\{\nbh(t)=\fbh\frac{\rhocdm(t)}{\mbh} \right\}\times \Gammabh(\mbh,\mchi,\tff,t)\,,
\een
where the individual rate $\Gammabh$ has been defined in \citeeq{eq:gammabh}, with a more effective expression in \citeeq{eq:gammabh_eff}, and where $\nbh$ and $\rhocdm$ stand for the PBH number density and for the homogeneous CDM mass density, respectively.

At this stage, since we have already discussed the dependencies in the main input parameters in the previous subsection, it is useful to make more explicit the time or redshift dependence of this global annihilation rate per unit volume. The redshift dependence of $\rhocdm$ is well-known, $\propto(1+z)^3$, so we can focus on $\Gammabh$. Though the time dependence was thoroughly reviewed before, in particular in the 3/2 and 9/4 slope cases, we shortly recall that if we neglect the residual time dependence in the factor $K_\gamma$, then according to \citeeq{eq:gammabh_eff}, we have:
\ben
\Gammabh(t)\propto \sigvav (\rcore(t))^3\left(\dfrac{\rhomax(t)}{\mchi}\right)^2 = (\rgamma)^3\left((1-\fbh)\rhogamma\right)^{3/\gamma}
\sigvav^{3/\gamma-1}\left(\frac{\Delta t}{\trec}\right)^{3/\gamma-2}\,,\nn\\
\een
where the time dependence naturally arises, and where $\trec$ refers to the recombination time, taken here as a reference time.
Further assuming that $t/\trec\simeq[(1+z)/(1+\zrec)]^{-3/2}$ (valid in matter domination) and identifying $\rhomaxrec=\rhomax(\trec)$ in the above expression, we finally get, after including all the missing terms:
\ben
\label{eq:gammabhz}
\Gammabh(z)\simeq \Gammabhrec \times \left(\frac{1+z}{1+\zrec}\right)^{3-9/(2\gamma)}\;,
\een
where $\Gammabhrec$, which generally depends on the fundamental parameters $\mchi$, $\mbh$, $\xkd$, and $\sigv$, is defined in the previous subsection -- see \citeeq{eq:gammabh_eff} for the general case, and \citeeq{eq:gammabh_32} or \citeeq{eq:gammabh_94} for the 3/2 or 9/4 slope cases, respectively.

Further noting that the PBH number density can be expressed as:
\ben
\nbh(z) = \fbh\frac{\rhocdm(z)}{\mbh} = \left\{\nbhrec\equiv\fbh \frac{\rhocdmrec}{\mbh}\right\}\left(\frac{1+z}{1+\zrec}\right)^{3}\;,
\een
where $\rhocdmrec=\rhocdm(\zrec)$, then we can rewrite the global annihilation rate density introduced in \citeeq{eq:gammabhtot} as:
\ben
\label{eq:gammabhtotz}
\frac{\dd \Gammabht(z)}{\dd V} \simeq 
\fbh \,\frac{\rhocdmrec}{\mbh}
\,\Gammabhrec\,\times
\begin{cases}
\displaystyle \left(\frac{1+z}{1+\zrec}\right)^{3\left(2-\frac{3}{2\gamma}\right)}\;\;\;&\text{if $\rcore<\rspike$}\\
\displaystyle \left(\frac{1+z}{1+\zrec}\right)^{6} & \text{otherwise}
\end{cases}\;,
\een
where the full redshift (or time) dependence is now explicit. The upper case is the most relevant for the parameter space considered in the present study, while the lower case only accounts for annihilation inside the spike radius when it gets smaller than the annihilation core, neglecting further contributions from the secondary infall extent to the spike. In practice, we mostly have a transition between two BH mass regimes characterized by $\gamma=3/2$ (light to moderate mass regime), for which the annihilation rate around BHs is $\propto (1+z)^3$, and by $\gamma=9/4$ (massive BHs), for which that rate becomes $\propto (1+z)^4$. The former regime scales like the evolution of the BH number density itself as it should, while an additional time dependence in the average number of annihilations in spikes affects the latter regime. We emphasize the striking difference with the annihilation rate of the smooth particle DM component, which scales like $\propto\rhocdm^2\propto (1+z)^6$ (for spikes, this scaling still arises when the core size exceeds that of the spike, as shown in the lower case above, but this behavior is mostly a consequence of neglecting the DM halo that should extend beyond the spike after secondary infall -- this specific situation would therefore be an incomplete description, but is not encountered in CMB analyses). The dependence on the other input model parameters are fully contained in $\Gammabhrec$, which was discussed earlier in \citesec{ssec:gamma_bh}. 

\subsection{Energy deposition and boost factor}
\label{ssec:deposition}

\begin{figure}[t!]
\centering
\includegraphics[width=\linewidth]{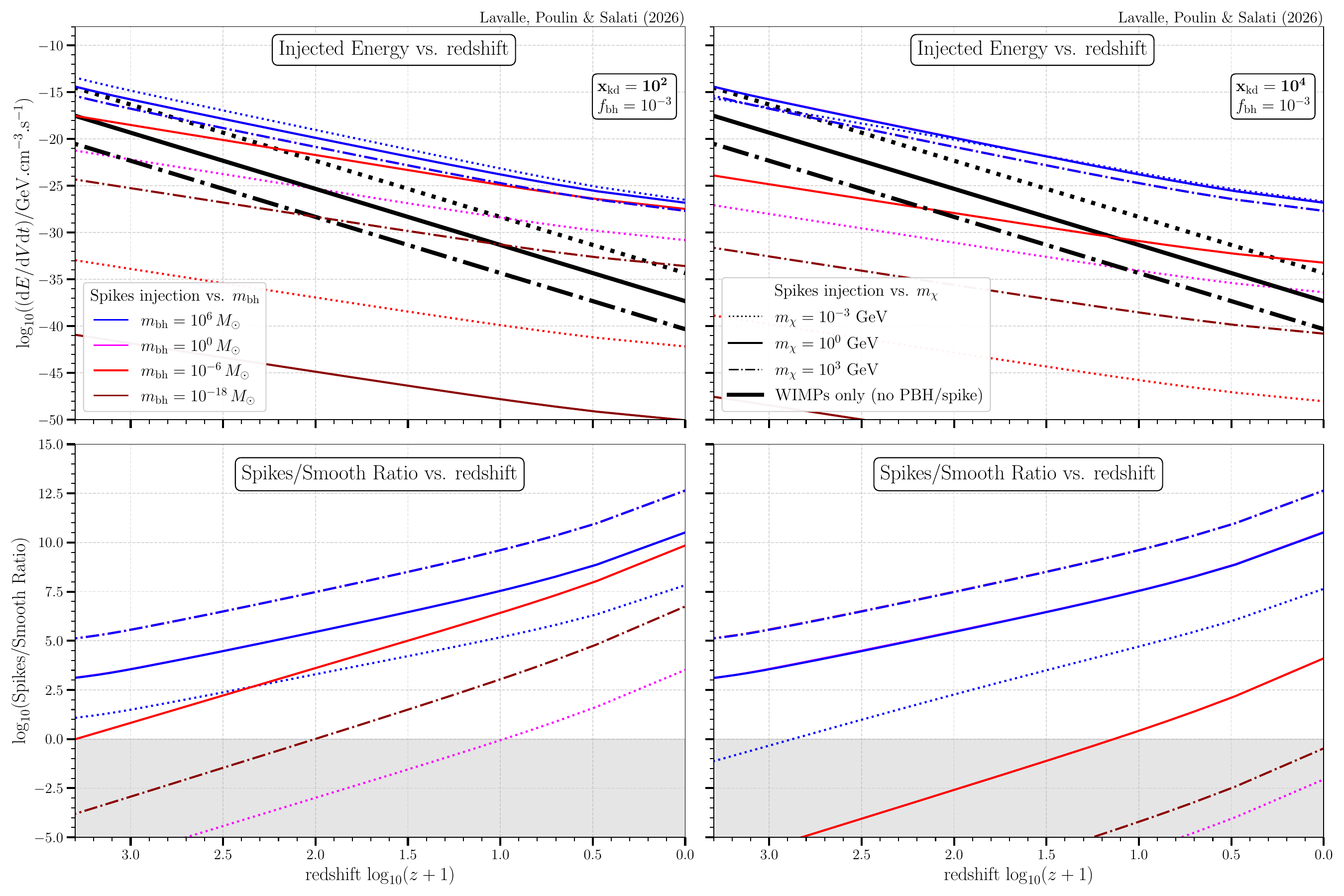} 
\caption{Injected energy density as a function of redshift for several monochromatic populations of BHs, with masses from $10^{-18}$ to $10^6$~$\Msun$ (line colors), assuming different WIMP masses from 1~MeV to 1~TeV (line styles). {\bf Top panels}: $\dd E(z)/\dd V \dd t$ for different configurations. {\bf Bottom panels}: Associated ratio with respect to the smooth DM component (without spikes). {\bf Left/right panels}: early/late kinetic decoupling. Note that the curves obtained for $\mbh\geq 1\,\Msun$ expectedly superimpose by virtue of \citeeq{eq:dEspikes_scaling}.}
\label{fig:dEdVdt}
\end{figure}

We can now fully define the rate of energy density injection on the spot, which is the main quantity entering the calculation of CMB constraints. First, let us assume that WIMP annihilation proceeds in the injection of a spectrum $\dd N/\dd E$ of standard model particles. Since the total energy injected per annihilation is $2\,\mchi$ (annihilation at rest), the particle spectrum must obey the local closure relation:
\ben
\int_0^{\mchi}\dd E \,E\,\frac{\dd N(E)}{\dd E} = 2\,\mchi\,.
\een
We further introduce a spectral injection source function $\dd {\cal Q}_{\rm spikes}$ as:
\ben
\label{eq:qsource}
\frac{\dd{\cal Q}_{\rm spikes}(E,z)}{\dd E} \equiv \frac{\dd \Gammabht(z)}{\dd V}\times \frac{\dd N(E)}{\dd E}\,,
\een
which characterizes the total energy injected by a population of spikes per units time and volume:
\ben
\dEspikes = \int_0^{\mchi}\dd E \,E\,\frac{\dd{\cal Q}_{\rm spikes}(E,z)}{\dd E} \simeq
2\,\mchi\,
\fbh \, \frac{\rhocdmrec}{\mbh}
\,\Gammabhrec\,\times
\left(\frac{1+z}{1+\zrec}\right)^{3\left(2-\frac{3}{2\gamma}\right)}
\;.\nn\\
\label{eq:dEdVdt_spikes}
\een
Here, we have only considered the cases for which spikes radii are larger than the annihilation cores -- see \citeeq{eq:gammabhtotz}. This expression, which obviously adds up to the contribution of the homogeneous particle DM component, will be at the basis of our CMB analysis. As for the homogeneous contribution itself, it reads:
\ben
\label{eq:dEwimps_fbh}
\dEwimps = (1-\fbh)^2\, \dEOwimps\,,
\een
which relates to the standard WIMP contribution in the absence of PBHs:
\ben
\label{eq:dEwimps}
\dEOwimps&\equiv& 2\,\mchi\,
\frac{\sigvav}{2} \,\left(\frac{\rhocdm(z)}{\mchi}\right)^2
= 
\frac{\sigvav}{\mchi}\rhocdm^2(z)\\
& = & 2.24\times 10^{-23} \,{\rm GeV/cm^3/s} \nn\\
&& \times
\left(\frac{\sigvav/(3\times 10^{-26}\,{\rm cm^3/s})}{\mchi/100\,{\rm GeV}}\right) \left(\frac{\Omega_{\rm cdm }h^2}{0.11933}\right)^2 \left( \frac{(1+z)}{601}\right)^6\,.
\nn
\een
The above numerical evaluation is given at a redshift of 600, where distortions in the CMB induced by DM annihilation are maximized \cite{FinkbeinerEtAl2012}.

The total rate of injected energy per volume is then given by:
\ben
\label{eq:dEtot}
\dEtot &=& \dEspikes + \dEwimps\\
&=&  \left(  1+{\cal B}_0(z) \right) \, \dEOwimps = {\cal B}(z) \, \dEOwimps\,,\nn
\een
where in the second line, we have introduced two complementary definitions of a boost factor that relates the total injected energy of the mixed PBH+WIMP scenario to that of the WIMP-only case. These boost factors can be easily integrated into numerical codes used in CMB data analyses, like the CLASS code \cite{Lesgourgues2011,BlasEtAl2011}. More precisely, they read:
\ben
\label{eq:boost}
{\cal B}(z) & \equiv & 
(1-\fbh)^2 + \dEspikes/\dEOwimps\\
&\simeq & (1-\fbh)^2 + 
2\dfrac{\mchi^2}{\sigvav}\dfrac{\fbh}{\rhocdmrec}\dfrac{\Gammabhrec}{\mbh}\left(\dfrac{(1+z)}{(1+\zrec)}\right)^{-9/2\gamma}\nn \\
{\cal B}_0(z) & \equiv & {\cal B}(z) - 1\,.\nn
\een
This expression characterizes the redshift-dependent departure from the homogeneous self-annihilating DM case, is specific to this mixed PBH+WIMP scenario, and depends on the spike profile slope $\gamma$ (on top of and related to all of the other fundamental parameters -- WIMP and BH masses, annihilation cross section, BH fraction, and kinetic decoupling temperature). We see that the boost factor increases with forward time (decreasing redshift) as it should, as the squared homogeneous density dilutes faster than the population of BHs. We will discuss the dependence on the other fundamental parameters of the scenario below (it is not explicit in the above form because $\Gammabhrec$ itself depends on these parameters).

An illustration of the rate of injected energy density is displayed in \citefig{fig:dEdVdt}, where the top panels show the redshift dependence for early (left panel) and late kinetic decoupling (right panel, respectively), and the bottom panels show the corresponding boost factors; a BH fraction of $0.1\%$ is assumed. The thick black curves in the top panels correspond to the WIMP-only case for three choices of particle masses, $\mchi= 10^{-3}$, 1, and $10^3$~GeV (dotted, solid, and dot-dashed curves, respectively). The other thin curves show the contribution of spikes with a color code related to the BH masses (from $10^{-18}$~$\Msun$ -- red -- to $10^6$~$\Msun$ -- blue). We see that the injection scales like $(1+z)^3$ for very light BHs, for which the number of annihilations within the extending core is constant with time, and $(1+z)^4$ for moderately massive to heavy BHs, all decreasing much more slowly with time than the smooth particle DM contribution $\propto (1+z)^6$. The amplitude of the spikes signal is almost independent of the BH mass for heavy BHs (beyond the breaking mass) because the mass dependence of the individual decay rate $\Gammabh\propto \mbh$ is compensated by the one in the number density $n_{\rm bh}\propto \rhocdm/\mbh$, while we have $\dd E/\dd V\dd t\propto \mbh^2$ in the light mass regime because $\Gammabh\propto \mbh^3$ in that case. The bottom panels show that the boost factor is accordingly much larger for spikes around heavy BHs [growing with time like $(1+z)^{-2}$], while it can drop below one for the lightest BHs surrounded by light WIMPs [though growing faster with time,  $\propto (1+z)^{-3}$].

It is interesting to derive the full dependence on the fundamental parameters of this mixed DM scenario ($\mchi,\sigvav,\xkd,\mbh,\fbh$) beside redshift only. These parameters are hidden in the expression of $\Gammabh$ in \citeeq{eq:dEdVdt_spikes}. By plugging in the results of \citeeq{eq:gammabh_32} and \citeeq{eq:gammabh_94} for the spike indices $\gamma$ of 3/2 (light BHs) and 9/4 (massive BHs), respectively, we obtain:
\ben
\label{eq:dEspikes_scaling}
\dEspikes \propto
\begin{cases}
    \fbh\,(1-\fbh)^2 \,\sigvav \,\xkd^{-3}\,\mchi^5\mbh^2 \; &\text{for}\;\gamma=3/2\;\text{(light BHs)}\\
    \fbh \,(1-\fbh)^{4/3}\,\,\sigvav^{1/3} \,\mchi^{-1/3}&\text{for}\;\gamma=9/4\;\text{(heavy BHs)}
\end{cases}\,.
\een
The absence of any BH mass dependence in this latter case (i.e. for BH masses greater than the breaking mass) is noteworthy. In particular, it explains why some of the curves of injected energy for the same (heavy enough) WIMP masses but for different BH masses above 1~$\Msun$ superimpose in \citefig{fig:dEdVdt} (and consequently those of the corresponding boost factors too --- except for the lightest WIMPs considered). Actually, a straight understanding of \citefig{fig:dEdVdt} is not so easy because the breaking mass, which positions the change of asymptotic regime from which the injection amplitude dependence on the WIMP mass gets reverted, does also depend on the WIMP mass among other parameters. The reader will strongly benefit from a careful cross-examination of \citefig{fig:gamma_bh_single} to make better sense of \citefig{fig:dEdVdt}.

\subsection{Extracting limits from the CMB data: the on-the-spot approximation}
\label{ssec:cmb_approx}

As a last step in this section, we can try to anticipate the order of magnitude of the CMB limits that will be derived from a complete data analysis in the next section. To do so, we can extrapolate the limits derived by the Planck Collaboration \cite{PlanckCollabEtAl2020a}. Assuming that the truly deposited energy can be estimated from the injected energy discussed above (at a given redshift of $\sim 600$) times an efficiency factor $\feff$ that depends on the annihilation final states -- the so-called on-the-spot approximation \cite{FinkbeinerEtAl2012} -- a 95\% confidence-level constraint was formulated in terms of an effective deposition factor in ref.~\cite{PlanckCollabEtAl2020a}:
\ben
p_{\rm ann}\equiv \feff \frac{\sigvav}{\mchi} < 
p_{\rm ann}^{\rm max} = 3.2\times 10^{-28}\,{\rm cm^3/s/GeV}\;.
\label{eq:pann}
\een
This limit is valid for a WIMP-DM universe, and relates directly to the expression of the standard WIMP injected energy given in \citeeq{eq:dEwimps}. By picking up the value of DM density at $z=600$, we can turn this into a limit on the injected energy as:
\ben
\dEspikesdep \simeq \feff \dEspikes \leq \dEdep^{\rm max} = p_{\rm ann}^{\rm max}\,\rhocdm^2(z=600)\,,
\een
where the subscript ``dep" stands for ``deposited" (to contrast with ``injected"). We stress that extrapolating the limit that way amounts to neglect the difference in redshift dependence of energy injection between homogeneous self-annihilating DM [$\propto (1+z)^6$] and spikes [$\propto (1+z)^4$ for heavy BHs], which should lead to some errors that we will try to estimate later. However, still following that assumption for the sake of the qualitative exercise, we can readily infer the way the fundamental parameters of our mixed DM scenario will be constrained by comparing the above limit with \citeeq{eq:dEspikes_scaling}. We see that above the breaking mass, \ie~in the 9/4-slope regime, the bound on the BH fraction will not depend on the BH mass. In contrast, below the breaking mass, the bound on the BH fraction will degrade $\propto \mbh^{-2}$ with decreasing BH mass.

In more operational language, the approximate limit reads:
\ben
\mchi\,\fbh\,\frac{\Gammabh^{(600)}}{\mbh}
\lesssim \frac{p_{\rm ann}^{\rm max}}{2\,f_{\rm eff}}\,\rhocdm^{(600)}\,,
\label{eq:approx_cmb_limit}
\een
where the superscript $(600)$ indicates that the related quantity is evaluated at redshift 600 --- given the way $p_{\rm ann}^{\rm max}$ is derived in standard CMB data analysis, it would barely make sense to use this relation at other redshifts, even though we know that the time dependence of energy injection differs from usual dark matter annihilation. By taking the scaling relations of \citeeq{eq:gammabh_32} and \citeeq{eq:gammabh_94} together with the more precise calculation results given in the text just below (at $z=600$), we get for light BHs (3/2-slope regime):
\ben
\fbh \, (1-\fbh)^2\,
 \left[\frac{\sigvav}{3\times 10^{-26}\,{\rm cm^3/s}}\right] \,
\left[\frac{\xkd}{10^4}\right]^{-3}\,
\left[\frac{\mchi}{100\,{\rm GeV}}\right]^5\,
\left[\frac{\mbh}{10^{-18}\,\Msun}\right]^2 \lesssim \frac{8.25\times 10^{12}}{\feff}\nn\\
\label{eq:rough_bound_32}
\een
and then for heavy BHs (9/4-slope regime):
\ben
\fbh\,(1-\fbh)^{4/3}\,
\left[\frac{\sigvav}{3\times 10^{-26}\,{\rm cm^3/s}}\right]^{1/3}\,
\left[\frac{\mchi}{100\,{\rm GeV}}\right]^{-1/3}
\lesssim \frac{5.65\times 10^{-9}}{\feff}\,.\nn\\
\label{eq:rough_bound_94}
\een
These estimates provide rough indications about the level of limits one can extract from CMB data, and define very clearly highly non-trivial dependencies in all of the relevant defining parameters. Assuming $f_{\rm eff}\sim 0.1$ \cite{SlatyerEtAl2009,FinkbeinerEtAl2012,Slatyer2016}, it is rather obvious that in the very light BH mass range, limits will be mostly irrelevant below $\mbh \sim 10^{-11}$~$\Msun$ (except for very heavy WIMPs $\gg 100$~GeV), which roughly defines the sensitivity threshold of the CMB data to this mixed scenario in terms of BH mass. Moreover, since (still neglecting the homogeneous particle DM component) the injected energy density scales like $\propto \fbh\, (1-\fbh)^\nu $, it increases with the fraction $\fbh$ up to a turning point defined by $\fbh = 1/(\nu+1)$, above which the signal gets suppressed with increasing BH fraction. This denotes a loss of constraining power as the BH fraction approaches unity, since WIMPs are no longer abundant enough to significantly accumulate around BHs and efficiently self-annihilate\footnote{This holds for a fixed annihilation cross section, but rigorously, if the latter derives from the freeze-out mechanism, it should consistently compensate for a factor of $(1-\fbh)^{-1}$ to accommodate the presence of BHs as a faction $\fbh$ of the dark matter. One can readily infer the resulting scaling relation, but the general picture keeps holding.}. In this case, only heavy PBHs can be constrained from the radiative transfer following baryonic accretion \cite{Ali-HaiemoudEtAl2017,PoulinEtAl2017}. In the reverse case where the PBH fraction gets vanishingly small, the only remaining CMB limits are those on the homogeneous WIMP component.

\begin{figure}[t!]
\centering
\includegraphics[width=0.49\linewidth]{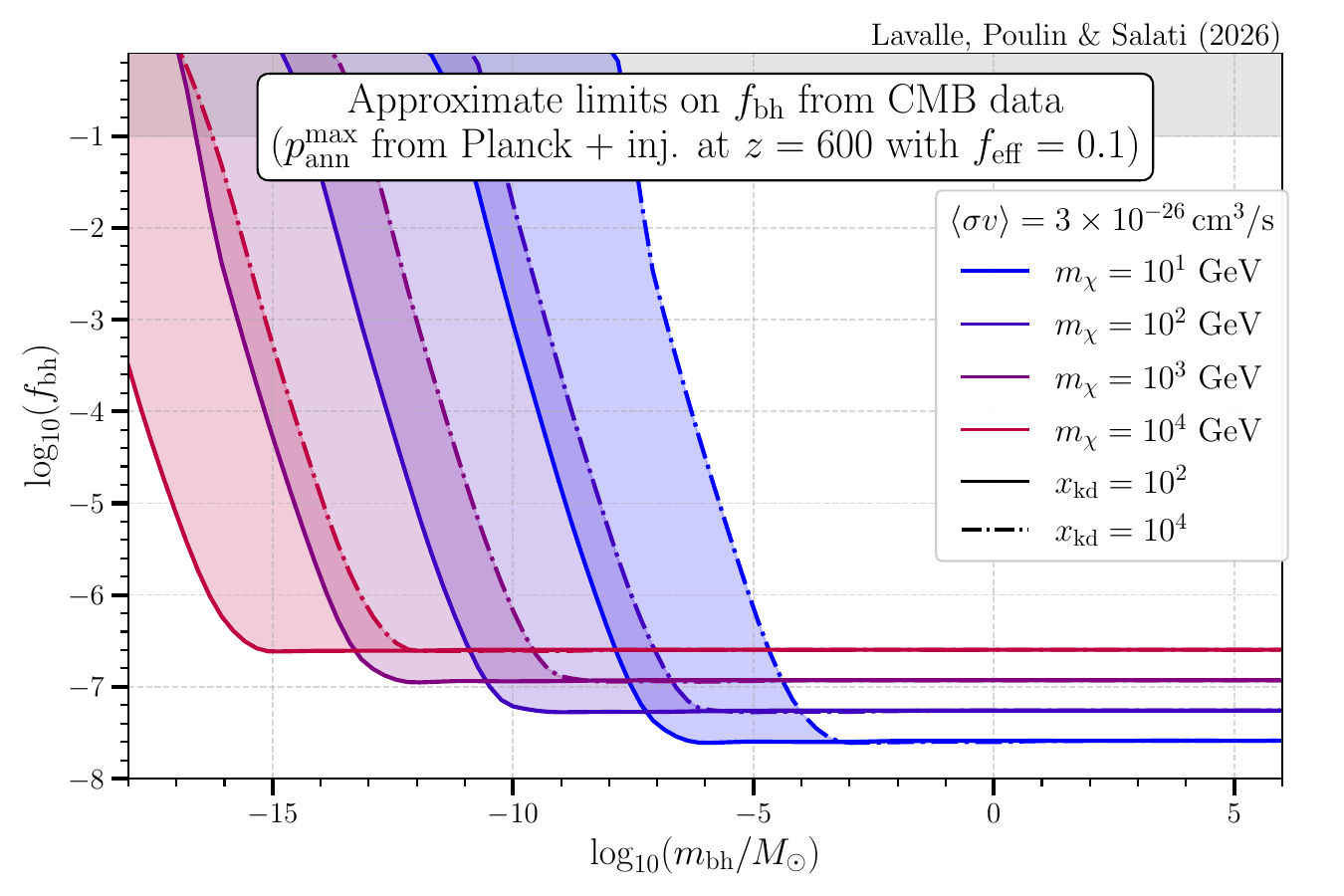} 
\includegraphics[width=0.49\linewidth]{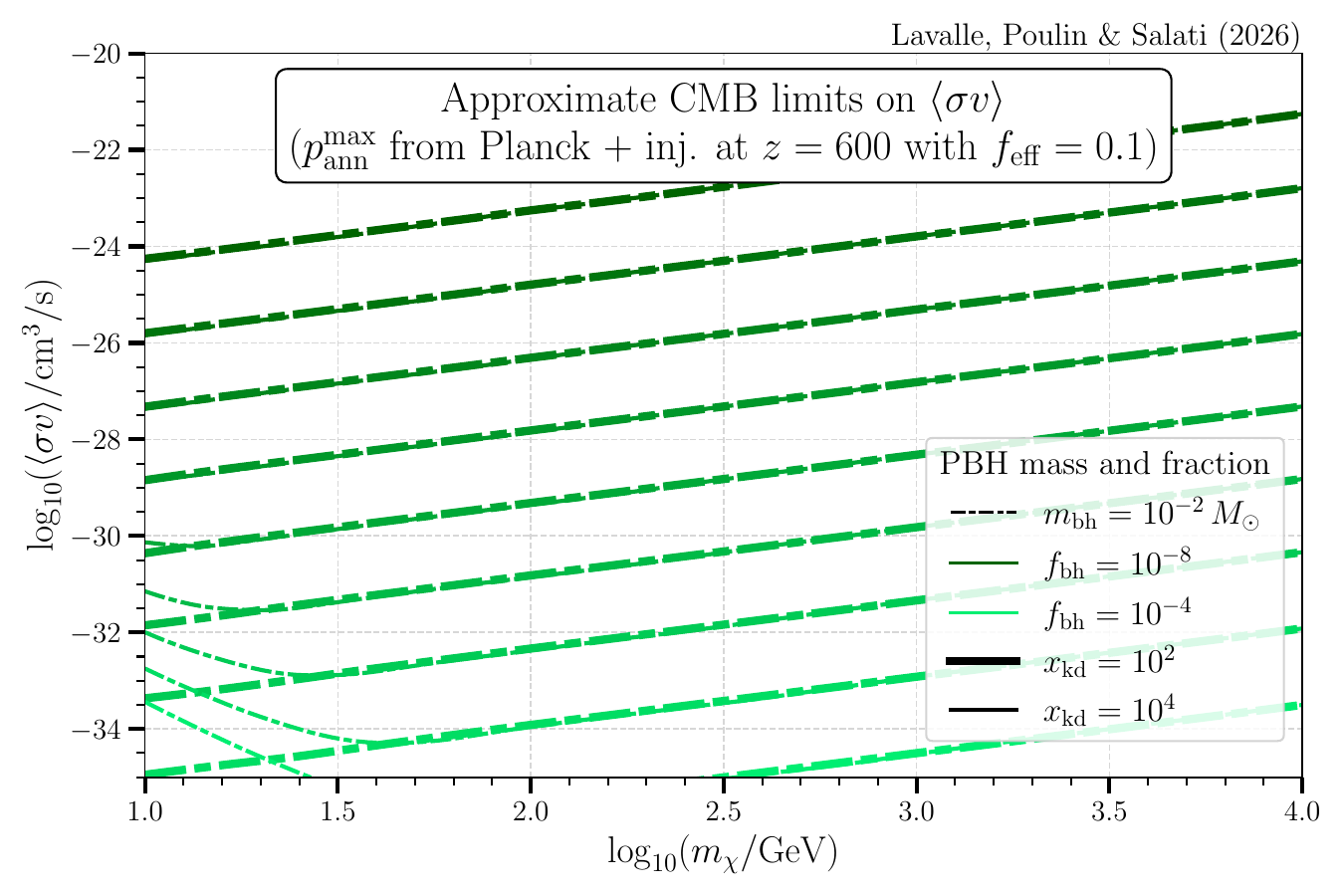} 
\caption{{\bf Left panel:} Approximate limits on the BH fraction as a function of their masses, assuming an extra DM component made of WIMPs of canonical annihilation cross section, and taking different WIMP masses. The on-the-spot approximation is used with a deposition efficiency of $f_{\rm eff}=0.1$, calculated at a redshift $z=600$. Colored bands show how the limit changes between two values for $\xkd$, $10^2$ (early), and $10^4$ (late kinetic decoupling). {\bf Right panel:} Approximate limits on the WIMP annihilation cross section ($s$-wave) assuming a population of PBHs of mass 0.01~$\Msun$, and relative abundance $\fbh$ ranging from $10^{-8}$ to larger values by steps of 0.5 dex (dark to light green). Thick (thin) curves correspond to early (late) kinetic decoupling, and are in most cases superimposed. The energy deposition is taken at $z=600$.}
\label{fig:cmb_limits_extrap}
\end{figure}

In \citefig{fig:cmb_limits_extrap}, we show the approximate limits on the mixed PBH-WIMP scenario derived by extrapolating the Planck results following the previous discussion. On the left panel, we show the expected constraints on the BH fraction $\fbh$ as a function of the BH mass $\mbh$ assuming WIMPs of different masses annihilating with the canonical cross section. At this stage, the annihilation channel is not important, and we only need to fix the effective energy deposition efficiency, $\feff=0.1$ here. We plotted the on-the-spot approximation results calculated at redshift $z=600$, even though we remind that the energy deposition history should in principle differ from the standard WIMP scenario. Indeed, since the global spikes decay rate scales more smoothly with redshift, $\overset{\sim}{\propto}(1+z)^4$, than traditional DM annihilation, $\propto (1+z)^6$, we expect the actual redshift peak of injection-deposition to be slightly lower than the classical value of $z\sim 600$ \cite{SlatyerEtAl2009,Slatyer2016}. \change{However, plugging in a lower effective redshift of injection would not be consistent with the definition of $\pann$, as discussed above, so we stick to the nominal value}. Overall, we see the two regimes characterized by \citeeq{eq:rough_bound_32} and \citeeq{eq:rough_bound_94}, for light (3/2 slope) and heavy (9/4 slope) BH spikes respectively, showing up quite clearly. In the heavy BH mass range, the constraint on the BH fraction does not depend on the BH mass, but does depend on the WIMP mass $\propto\mchi^{1/3}$. A striking feature is the reach of the constraints in this heavy BH mass, plateau regime, which goes as low as $\fbh\lesssim 10^{-7}-10^{-8}$. On the contrary, as BHs get lighter and lighter and cross the breaking mass, the constraints on $\fbh$ relax with decreasing BH mass $\propto \mbh^{-2}$. Finally, though not visible from the plot, we emphasize that in the plateau regime, the dependence on the cross section is much weaker than in the standard annihilation case, $\propto \sigvav^{1/3}$, which indicates that even suppressed annihilation cross sections are expected to lead to very constraining bounds \cite{ChandaEtAl2025}, which has actually been known for some time \cite{LackiEtAl2010,Eroshenko2016} though on less accurate grounds.

In the right panel of \citefig{fig:cmb_limits_extrap}, we show the reverse view by constraining the annihilation cross section as a function of WIMP mass, for populations of BHs of mass $\mbh=10^{-2}$~$\Msun$ with relative abundances $\fbh$ ranging from $10^{-8}$ (dark green) up to larger values (lighter green curves, respectively) -- with steps of 0.5 dex in $\fbh$ between each curve (what lies above a curve corresponds to an excluded region). This choice of BH mass is such that we sit in the heavy mass regime above the breaking mass, and would correspond to a point located on the plateau regime of the left panel's limits on $\fbh$. We again consider the injection-deposition to occur at $z=600$. Thick (thin) curves indicate early (late) kinetic decoupling, and are mostly superimposed. A small difference in these two regimes is seen at large BH fraction (hence small allowed annihilation cross section) and light WIMP mass. This comes from the fact that the breaking mass [see \citeeq{eq:mbreak}] increases with decreasing cross section and WIMP mass, and increasing $\xkd$, so that it becomes larger than $10^{-2}$~$\Msun$ in this corner of the parameter space, signaling a transition from the 9/4- to the 3/2-slope regime. Otherwise, it is again striking to see that the limits follow exactly the power law $\sigvav^{\rm max}\propto\mchi$, are mostly independent of $\xkd$, and is strongly sensitive to the BH fraction and deposition efficiency, since scaling like $\overset{\sim}{\propto}(\fbh\,f_{\rm eff})^{-3}$. All this is fully predicted by the approximate result summarized in \citeeq{eq:rough_bound_94}. However, we emphasize that for BH masses significantly lower than the breaking mass given in \citeeq{eq:mbreak}, the constraining power of BHs upon WIMPs is essentially lost (as is the case in the asteroid mass range).\\

Before moving to the full statistical analysis of the CMB data, several comments are in order.

First, in the derivation of the approximate limits in \citefig{fig:cmb_limits_extrap}, we have assumed a deposition of energy on the spot, with a fixed efficiency $\feff$, independent of the WIMP mass. However, as is well established \cite{SlatyerEtAl2009,Slatyer2016a}, the energy deposition following injection from DM annihilation, and its efficiency to heat or ionize the medium, is a complex process characterized by a non-trivial history that strongly depends on energy. Therefore, the integrated energy deposition after recombination will be different for the annihilation of particles of different masses, not to mention the effects of the annihilation channels. Consequently, the very well ordered hierarchy linked to the WIMP mass observed in the approximate limits on $\fbh$ in \citefig{fig:cmb_limits_extrap}, in the plateau regime of the left panel, or the power-law behavior in limits on $\sigvav$ in the right panel, will actually be blurred and distorted by the actual energy deposition history in a more involved analysis; the relative amplitudes of the approximate limits will also be affected. We will see that more clearly in the next section.

Second, so far, we did not discuss a well-known physical effect which could lead to disrupting forming or formed spikes: early PBH binaries \cite{NakamuraEtAl1997,Ali-HaiemoudEtAl2017a,SasakiEtAl2018,RaidalEtAl2024}. These binaries could form already in radiation domination (or later), and prevent the formation of (or destroy) dark matter spikes; they have been studied in the context of predictions of the PBH merger rate, and stand as potentially copious sources of GWs. What is relevant here is truly the fraction of BHs that experienced strong tidal forces along their history, irrespective of merging which is of interest only for gravitational wave detection. We can very roughly estimate the number of early binaries as follows. First, the approximate average separation between two BHs is given by:
\ben
\lambda_{\rm bh}(\fbh,\mbh,z) &\equiv& n_{\rm bh}^{-1/3}\\
&\approx & 0.92\,\text{pc}\,\left(\frac{f_\mathrm{bh}}{10^{-3}}\right)^{-1/3}\,\left(\frac{\Omega_\mathrm{dm}\,h^2}{0.12}\right)^{-1/3} \,\left(\frac{M_\mathrm{bh}}{1\,M_\odot}\right)^{1/3}  \left(\frac{(1+z)}{3401}\right)^{-1}\,,\nn
\een
where the numerical estimate is done at equivalence. This has to be compared with the BH radius of influence at equivalence given in \citeeq{eq:rtspike}, which can be done by inspecting the ratio:
\ben
\label{eq:alpha_lonely}
\alpha_{\rm lonely}\equiv \frac{\lambda_{\rm bh}}{\rinfl}\Bigg|_{\rm eq}
\approx 33.07 \,\left(\frac{f_\mathrm{bh}}{10^{-3}}\right)^{-1/3}\,. 
\een
While $\alpha_{\rm lonely}\gg 1$, the formation of PBH binaries (and then the potential disruption of forming spikes) can safely be neglected during radiation domination. We see that this ratio does not depend on the BH mass, only on the relative BH abundance $\fbh$. Earlier in time or deeper in radiation domination, $\alpha_{\rm lonely}\propto a^{-1/3} \propto (1+z)^{1/3}$, so the evaluation at equivalence is a good proxy. Therefore, even though several BHs could be contained in the causal horizon, early disruption of spikes due to the formation of binaries can essentially be neglected except if $\fbh\sim 1$. This is consistent with more precise estimates \cite{RaidalEtAl2024}, since only a small fraction of PBHs in binaries is enough to strongly constrain their abundance beyond the solar mass range. In any case, even though a small correction to our calculation should in principle be necessary to exclude the contribution of binaries, it is expected to be small. Including it would actually add another time dependence reflecting the time it takes for BHs to fall onto one another, which depends on their distances to their closest neighbors and on torques generated by the next-to-closest ones or by forming DM halos. This goes beyond the scope of this paper.

\section{CMB constraints from a complete statistical analysis}
\label{sec:cmb}

\subsection{Analyses setup}
In this section, we present the complete statistical analysis carried out with the Planck CMB data \cite{PlanckCollabEtAl2020a}. To proceed, we implement our calculation of the PBH spikes injection rate presented earlier in the CLASS code (the ExoCLASS branch \cite{StoeckerEtAl2018}). More specifically, we modify the DarkAges module in order to account for this new type of energy injection --- this module already included the WIMP annihilation spectra from ref.~\cite{CirelliEtAl2011}, and the injected-to-deposited energy transfer functions developed in refs.~\cite{SlatyerEtAl2009,Slatyer2016a} for all relevant annihilation channels. 
DarkAges integrates the complex energy deposition history, accounting for the time it takes for the injected energy to heat, excite, or ionize the ambient medium. 
The spikes injection rate itself is tabulated after sampling the full numerical and accurate calculation of $\Gammabh$ of single spikes over a 5-dimensional grid featuring parameters $\mbh$, $\mchi$, $\xkd$, $\sigvav$, and the redshift.\footnote{We actually trade $\sigvav$ and redshift for $\rhomax^0$, which allows us to get the combination of both parameters in terms of a single variable, see \citeeq{eq:rhomax_exact} --- this reduces the dimensionality to 4 effective parametric dimensions.} This multidimensional table is interpolated over with a dedicated Python module, inside which the explicit (and non-trivial) dependence on $\fbh$ is further implemented [see \citeeq{eq:dEspikes_scaling}].

We perform Monte-Carlo Markov Chains (MCMC) analyses using MontePython-v3 \cite{AudrenEtAl2013,BrinckmannEtAl2019} and the Metropolis-Hastings algorithm.
We consider data combination including the \textit{Planck} 2018  high-$\ell$ TTTEEE \texttt{Plik} likelihood, the low-$\ell$ TT and EE likelihoods, as well as the public release 3 (PR3) lensing likelihood. We further include uncalibrated luminosity distances to supernovae of type 1a (SN1a) compiled in the \textit{PantheonPlus} catalog~\cite{Brout:2022vxf}, as well as baryonic acoustic oscillation measurements from the dark energy spectroscopic instrument (DESI) data release 2~\cite{DESI:2025zgx}.
We consider the six following $\Lambda$CDM parameters: the reduced baryon density $\omega_b$, the reduced cold dark matter density $\omega_{\rm cdm}$, the expansion rate today $H_0$, the primordial power spectrum amplitude $A_s$ and tilt $n_s$, and the optical depth to reionization $\tau_{\rm reio}$ (reduced densities are defined as $\omega_i\equiv \Omega_i h^2$, where $i$ represents any species). We impose large uninformative flat priors on those parameters, and make use of a Choleski decomposition to handle the large number of nuisance parameters. 
We further consider one massive and two massless neutrinos, with $\sum m_\nu=0.06$ eV as is conventional. From this up-to-date analysis setup, we can already rederive a 95\%-confidence-level (CL) limit on the standard self-annihilating WIMP dark matter scenario (no PBHs), which we can again express in terms of a maximal effective deposition factor:
\ben
\ptannmax = 2.94 \times 10^{-28}\,{\rm cm^3/s/GeV}\,,
\label{eq:new_pann}
\een
very close to the value formerly obtained by the Planck Collaboration and given in \citeeq{eq:pann}. The difference is due to the inclusion of the new DESI BAO data, which has  slightly strengthened the bound on models with electromagnetic energy injection that delay recombination (DESI rather favors earlier recombination, see e.g. Refs.~\cite{Mirpoorian:2025rfp,Lynch:2024hzh,SPT-3G:2025vyw,Chaussidon:2025npr}).

 
We consider several mixed PBHs and WIMPs scenarios. First, we assume that WIMPs exist, and derive constraints on the BH fraction as a function of BH mass.
 For definiteness, we specialize our analysis to WIMP configurations that are still immune to current indirect detection constraints (see for instance the compilation in Fig.~6.14 of ref.~\cite{CirelliEtAl2024}) ---  we remind that for \eg~fermionic WIMPs, $s$-wave annihilation processes mediated by, for instance, pseudo-scalar mediators, are essentially unconstrained by direct detection searches \cite{ArcadiEtAl2025}. To this aim, we pick up two masses, $\mchi=200$~GeV and 1~TeV, which are observationally allowed and within reach of current and future experiments, and  two different kinetic decoupling times set by $\xkd = 100$ (early) and $10^4$ (late decoupling). We consider $s$-wave annihilations into $b\bar b$ and $\tau^+\tau^-$ as generically representative of most WIMP scenarios.
 Second, we assume that BHs exist, and derive limits on the annihilation cross section $\sigvav$ as function of WIMP mass $\mchi$ (above 5~GeV), for the two different kinetic decoupling times and annhihilation channels specified just above. 
 In that second configuration, we fix the BH mass to $10^{-2}M_\odot$ and relative abundance to $\fbh=10^{-6}$, \ie~below current observational limits for PBHs but not far from the reach of GW experiments. In particular, the choice of a sub-solar mass ensures that an hypothetical discovery would point to a primordial origin.
 Importantly, in both cases we perform MCMC analyses by fixing the mass (either of BH or WIMP) and varying the corresponding remaining parameter ($\log_{10}f_{\rm BH}$ or $\log_{10}(\sigvav/({\rm cm^3/s})$) within large flat priors. The use of logarithmic priors is warranted given the large dynamical range of the parameters explored by the MCMC.
 We scan over a grid of ten masses equally spaced in $\log_{10}(m_{\rm PBH}/M_\odot)\in[-18,-4]$ or $\log_{10}(m_\chi/{\rm GeV})\in[0.7,4]$.
 We have indeed found that letting the PBH or WIMP mass free to vary prevent proper convergence of the MCMC.


\subsection{Constraints on $\{f_{\rm BH},m_{\rm BH}\}$}
\label{ssec:fbh}

\begin{figure}[t!]
\centering
\includegraphics[width=0.49\linewidth]{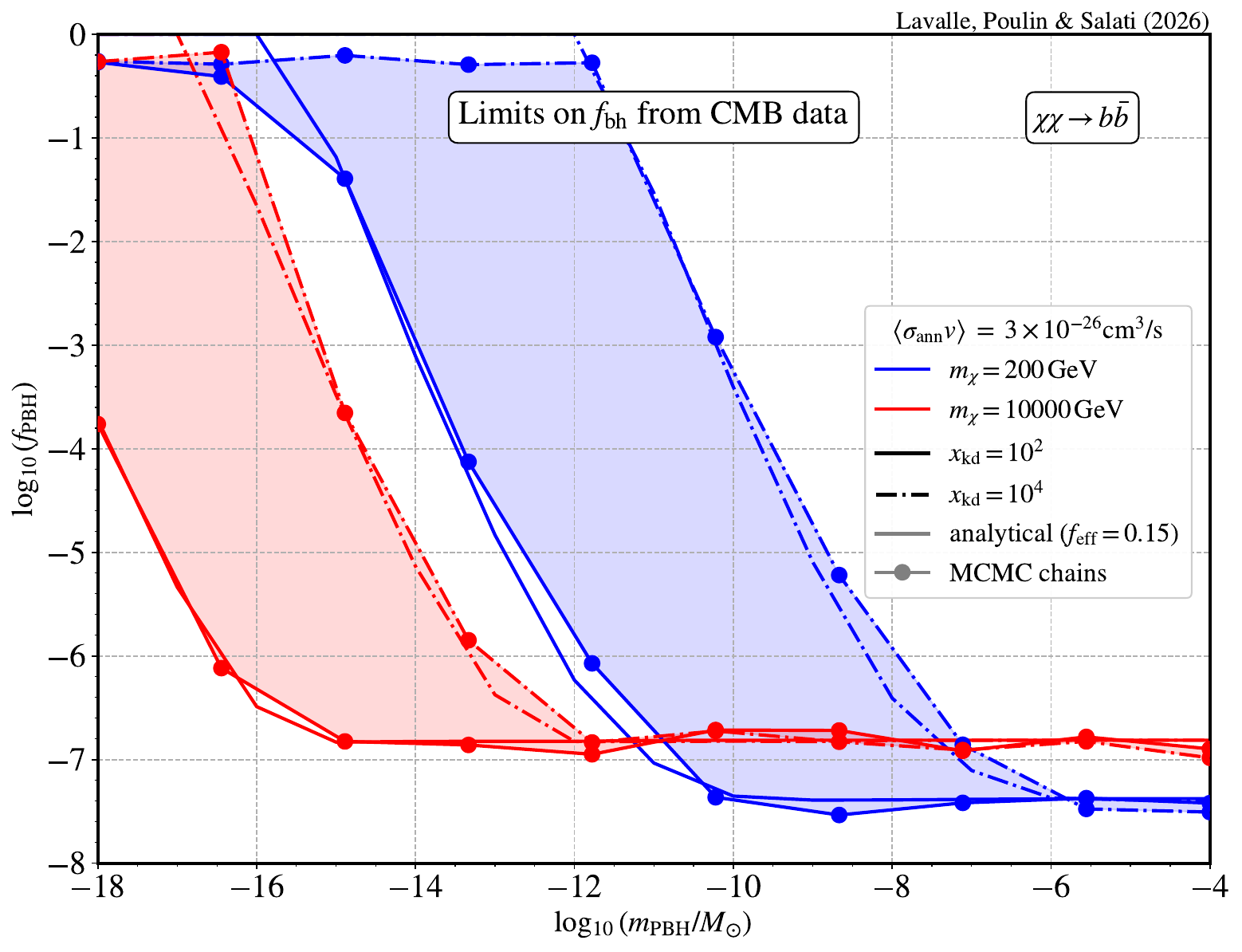} 
\includegraphics[width=0.49\linewidth]{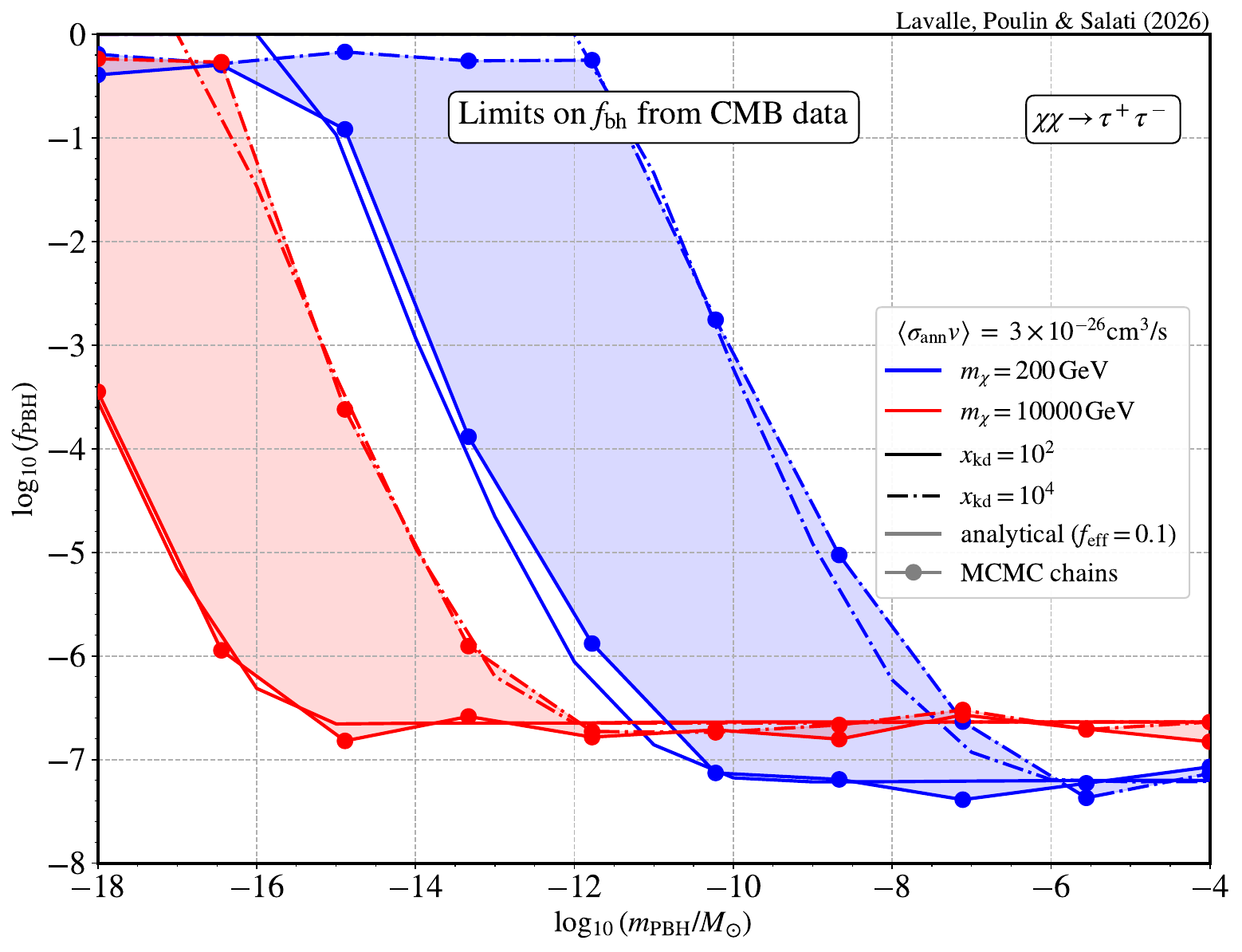} 
\includegraphics[width=0.49\linewidth]{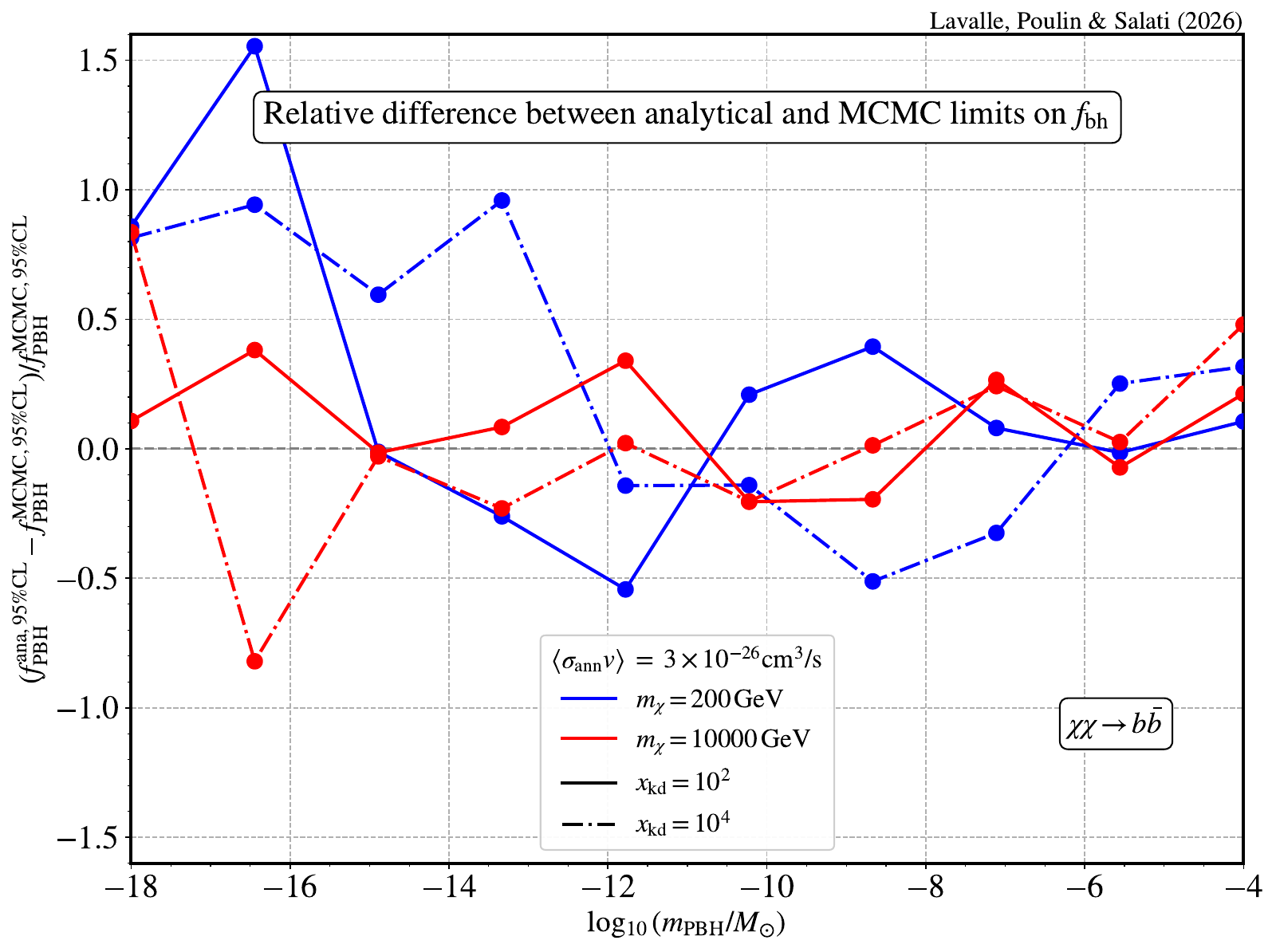} 
\includegraphics[width=0.49\linewidth]{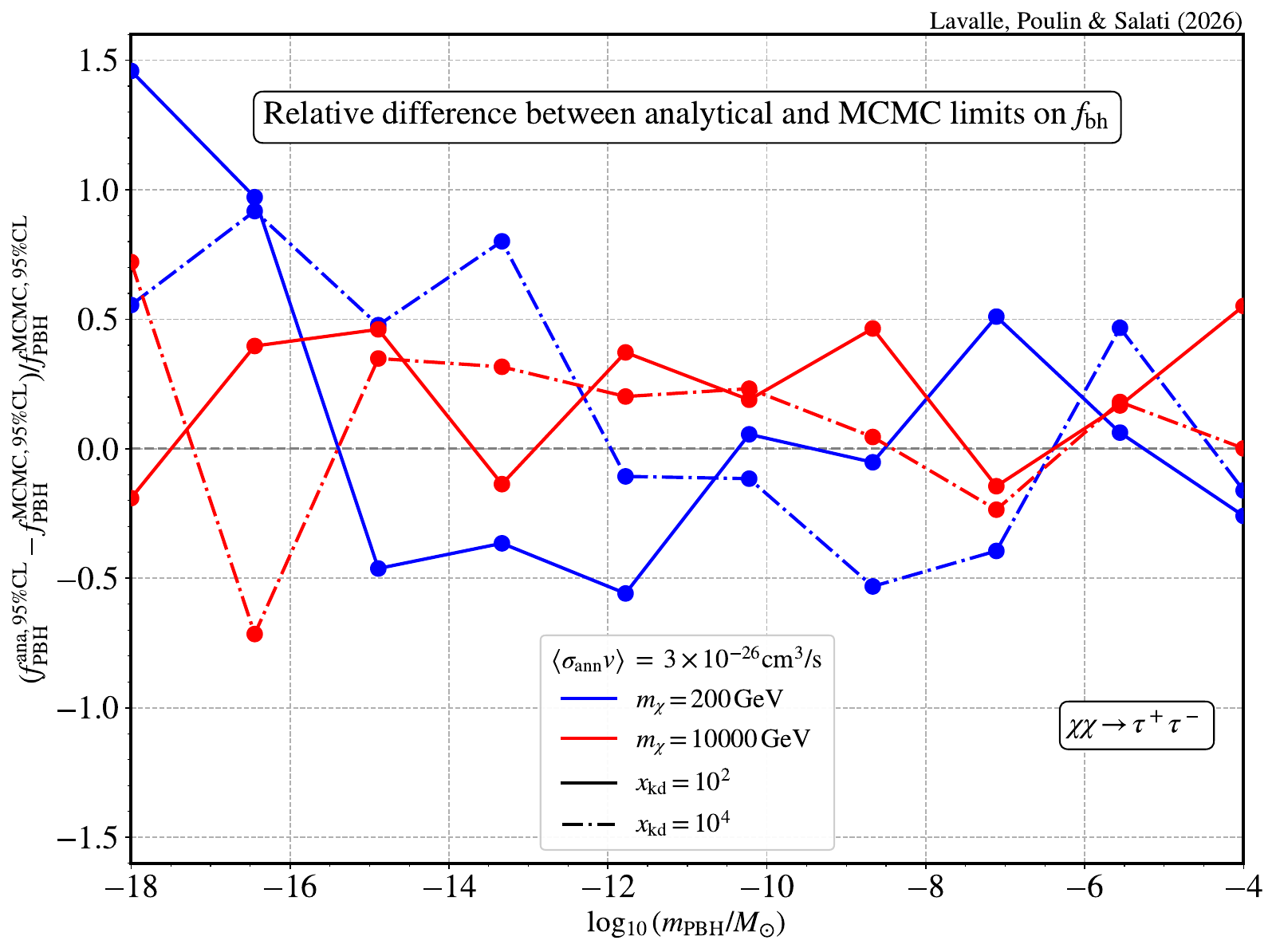} 
\caption{{\bf Top panels}: Full CMB+BAO+SN1a limits obtained on the BH fraction $f_{\rm PBH}$ as a function of their mass $m_{\rm PBH}$, assuming $\sigvav = 3\times 10^{-26}$ cm$^3$/s, two different WIMP masses of $m_\chi=200$ GeV (blue) and $10$ TeV (red), and a $b\bar b$  (left panel) or $\tau^+\tau^-$ (right panel) annihilation channel. The shaded bands show how the limits change when varying $\xkd$ from $10^2$ to $10^4$. {\bf Bottom panels}: corresponding relative differences between semi-analytical predictions and MCMC results.}
\label{fig:cmb_limits_on_fbh}
\end{figure}
We present our 95\%-CL limits on the fraction of DM in PBHs as a function of their mass in \citefig{fig:cmb_limits_on_fbh} (in the monochromatic limit), which mimic our approximate predictions in \citefig{fig:cmb_limits_extrap} (left panel); in this case, we assume the rest of DM to be made of WIMPs with the canonical annihilation cross section -- the left (right) panels of \citefig{fig:cmb_limits_on_fbh} show the results obtained assuming WIMP annihilations into $b\bar{b}$ ($\tau^+\tau^-$, respectively). We consider both early and late kinetic decouplings, which are respectively displayed as plain and dot-dashed curves, for the two different WIMP masses of 200~GeV (blue curves) and 10~TeV (red curves). In the two top panels, we compare the full MCMC limits (curves with solid circles) with semi-analytical predictions tuned to the current analysis (curves without circles). The latter limits make use of \citeeq{eq:approx_cmb_limit}, but we consistently plug in the effective deposition factor given in \citeeq{eq:new_pann}; furthermore, we take $f_{\rm eff}=0.15$ for the $b\bar{b}$ channel, and 0.1 for the $\tau^+\tau^-$ channel (values consistent with the ones recommended in ref.~\cite{Slatyer2016a}, but set a posteriori to better match the MCMC results). The relative differences between the semi-analytical predictions and the MCMC results are displayed in the bottom panels. \new{More details about the MCMC posterior distributions and correlations among parameters are given in the appendix, in \citefig{fig:post_fbh_18} and \citefig{fig:post_fbh_4}, only for illustration.}

First, we see that when using the injection-deposition effective parameters given just above, the simplified semi-analytical derivation of limits matches remarkably well with the MCMC results (see the bottom panels for the relative differences). The plateau regime is almost fully recovered by the MCMC, with small variations mostly due to statistical nature of parameter exploration, consistent with the requested CL. The scaling with the WIMP mass $\propto\mchi^{1/3}$ is nicely recovered, with more stringent limits for lighter WIMPs, though in smaller ranges of BH mass (due to the relative increase of the breaking mass). This confirms the potential strength of the limit on $\fbh$ above the asteroid mass range, should WIMPs be discovered someday --- the reach is around $\fbh\sim 10^{-7}$, and more stringent with $b\bar{b}$ than with $\tau^+\tau^-$ (obviously, light charged leptonic annihilation channels would significantly strengthen the limits as is well known \cite{Slatyer2016}). Furthermore, we note that the matching between the approximate prediction and the MCMC is within 50\% in the plateau, heavy BH mass regime (9/4 spike slope), but degrades in the light mass regime (3/2 slope). Although surprising owing to the significant level of approximation used to derive the semi-analytical limits, this still means that realistic and precise limits cannot really be obtained without a full statistical data analysis, and that quick estimates such as the one introduced in the previous section can only be taken as order-of-magnitude indications (indeed, we did set $\feff$ by hand a posteriori). Anyway, to summarize, our MCMC limit can be roughly captured by the following formula (annihilation into $b\bar{b}$):
\ben
\fbh^{\rm max} \approx 7\times 10^{-8}\,
\left(\dfrac{\feff}{0.15}\right)^{-1}
\left( \dfrac{\mchi}{200\,{\rm GeV}}\right)^{1/3}
\left( \dfrac{\sigvav}{3\times 10^{-26}{\rm cm^3/s}}\right)^{-1/3}\,.
\label{eq:approx_fbhmax}
\een
This quick formula, deriving from \citeeq{eq:rough_bound_94}, is valid above the breaking BH mass defined in \citeeq{eq:mbreak}, \ie~essentially above the asteroid mass range (9/4-slope regime). In that domain, the limit is fully independent of the BH mass.

These results can be compared with the study performed in ref.~\cite{GinesEtAl2022}, which is to our knowledge the only one in which a similar statistical analysis of the CMB data was carried out. The main difference with that study is that instead of relying on very approximated density profiles for the spikes as these authors did, we performed accurate calculations capturing all of the details of spikes' decays, which depend on all of the main input parameters: $\mbh$, $\fbh$, $\mchi$, $\sigvav$, and $\xkd$ (not to mention the cosmological parameters, which are also free parameters in the MCMC analysis). Therefore, if the analysis of ref.~\cite{GinesEtAl2022} in terms of a 3/2 or 9/4 power-law spike profile based on arguments similar to what was substantially more detailed in \citesec{sec:recipes} ends up with similar scaling relations for instance in terms of $\mbh$, the final results are matched only within an order of magnitude. In particular, the position of the PBH breaking mass, where the transition between the 3/2 and 9/4 profile index occurs, can be shifted significantly if we adopt a similar assumption for kinetic decoupling. However, on a qualitative perspective, the trends found are consistent, as one could have naively expected once two spike profile indices come into play --- we note that a log-normal mass function for PBHs was also considered in ref.~\cite{GinesEtAl2022}, which we keep for a future work. The challenging part we have addressed here was really to implement accurate calculations of spikes' decays within an MCMC study in a tractable manner. Complementary to this accurate implementation, note that we have also proposed more elaborate analytical approximations in \paperII\ that should allow anyone to calculate spike decay rates to similar precision, and which are probably easier to integrate into a CMB code.

Another study in ref.~\cite{KadotaEtAl2021} also considered CMB limits, but in an even more approximate way (both in terms of spike description and of the treatment on energy deposition pre- and post-recombination). That study focused on WIMPs annihilating through $p$-wave (velocity-dependent) processes. We keep the discussion of velocity-dependent annihilation to a further dedicated study.

\subsection{Constraints on $\{m_\chi,\sigvav\}$}
\label{ssec:sigv}
\begin{figure}[t!]
\centering
\includegraphics[width=0.49\linewidth]{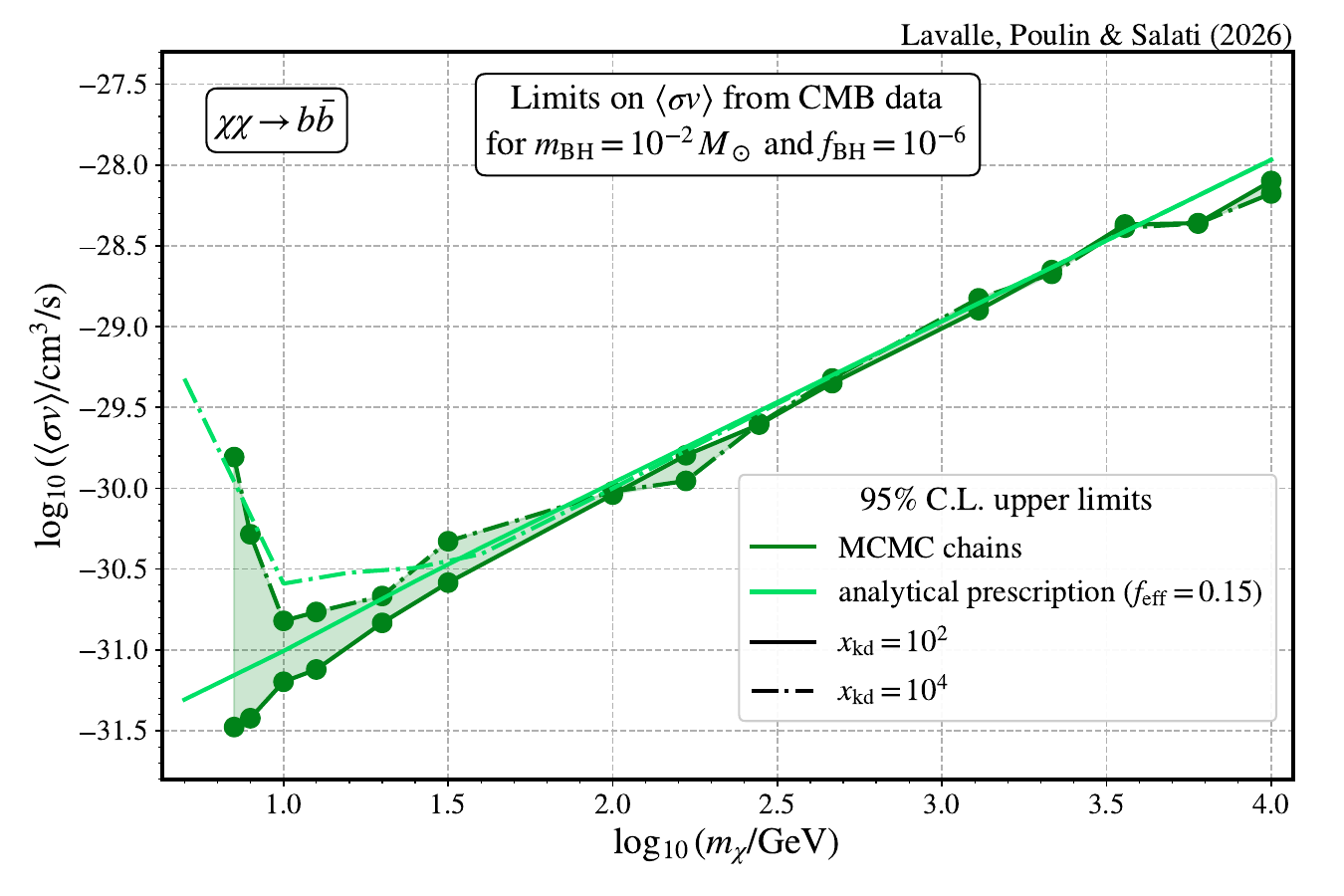} 
\includegraphics[width=0.49\linewidth]{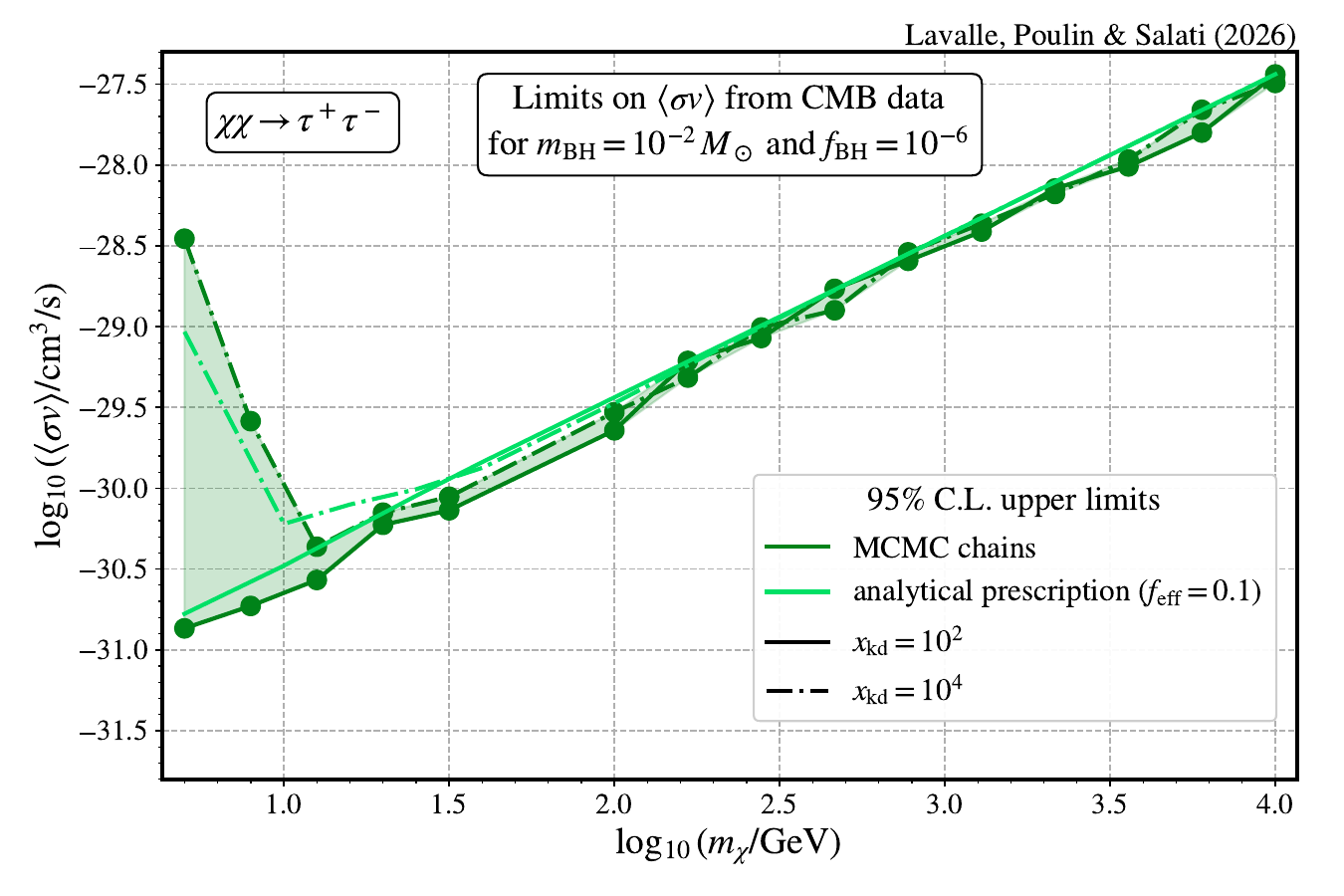} 
\includegraphics[width=0.49\linewidth]{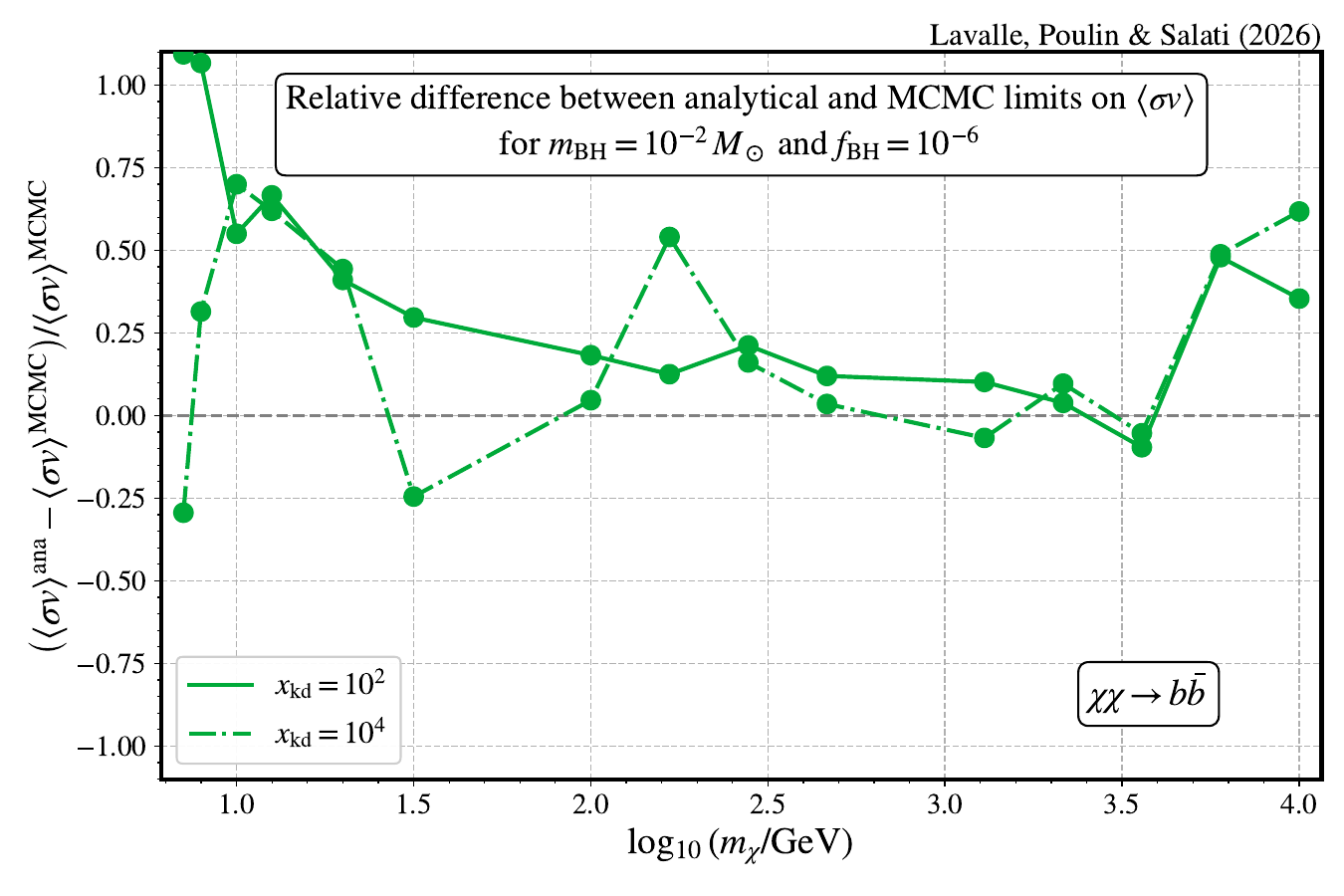} 
\includegraphics[width=0.49\linewidth]{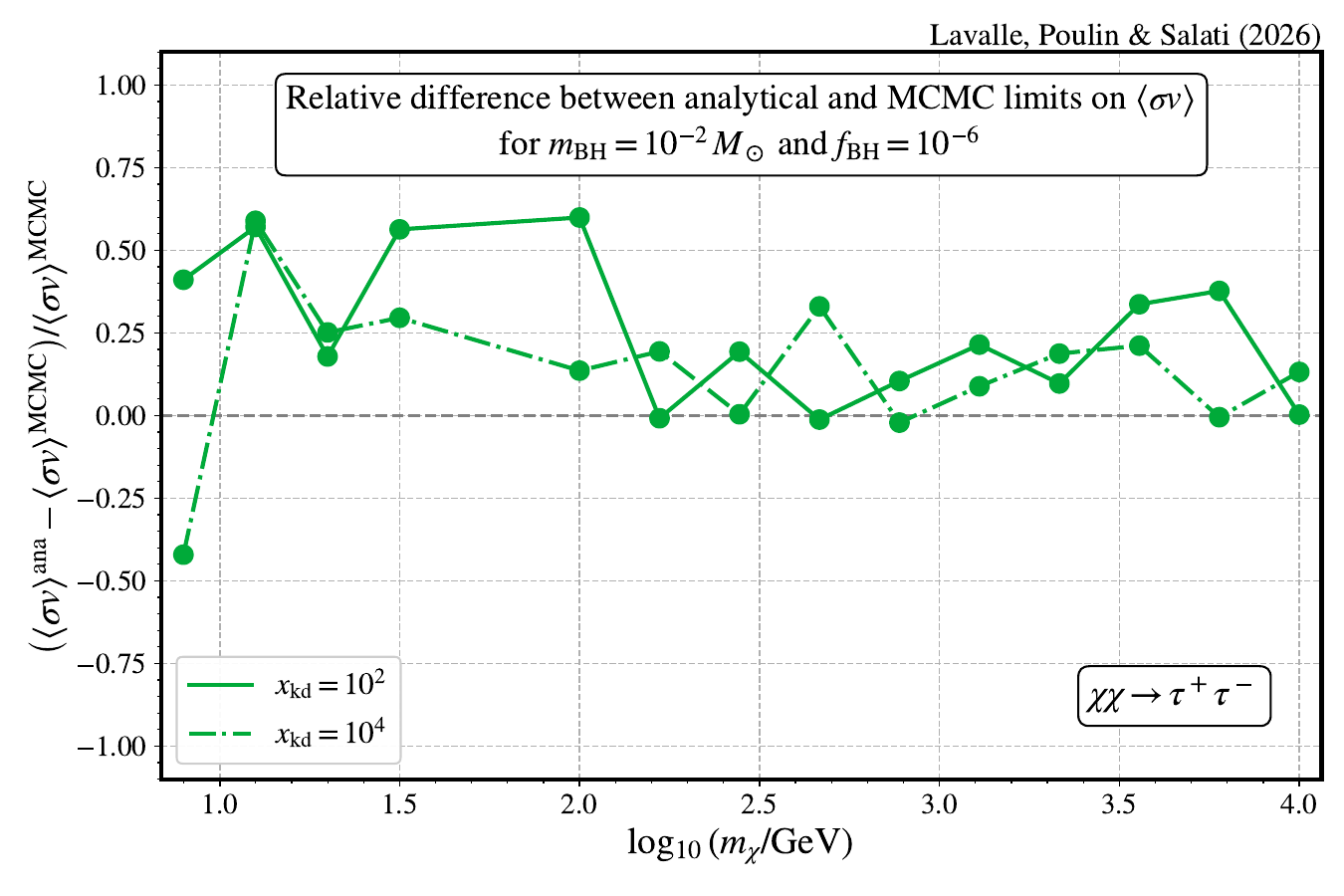} 

\caption{{\bf Top panels}: Full CMB+BAO+SN1a limits obtained on the DM annihilation cross section $\sigvav$ as a function of WIMP mass, assuming PBHs of masses 0.01~$\Msun$ and fraction $10^{-6}$, and a $b\bar b$  (left panel) or $\tau^+\tau^-$ (right panel) annihilation channel. The green bands show how the limits change when varying $\xkd$ from $10^2$ to $10^4$. {\bf Bottom panels}: corresponding relative differences between semi-analytical predictions and MCMC results.
}
\label{fig:cmb_limits_on_sigv}
\end{figure}

Our limits on the annihilation cross section as a function of WIMP mass, assuming a population of PBHs specified by their mass and fraction, are shown in \citefig{fig:cmb_limits_on_sigv}, which can be compared with the right panel plot of \citefig{fig:cmb_limits_extrap}. We consider a monochromatic mass distribution with $\mbh = 10^{-2}\,\Msun$, \ie~in the subsolar mass range, together with a tiny fraction of $\fbh = 10^{-6}$ --- these parameters imply that we are in the 9/4-slope regime. We further adopt the effective injection-deposition parameters already described in \citesec{ssec:fbh}. The left (right) panels show our results assuming a $b\bar{b}$ ($\tau^+\tau^-$) annihilation channel. In both cases, early (late) kinetic decoupling is represented with plain (dot-dashed) curves --- the shaded regions represent the spread induced by considering either early or late decoupling. MCMC results are displayed with dark green curves joining solid circles, while our semi-analytical predictions are shown as light green curves (without circles). As in the previous subsection, the matching between the predictions and the full MCMC results is remarkable --- the top panels show the full results, while the bottom ones show the relative differences between the semi-analytical limits and the MCMC results. The expected scaling relations given in \citeeq{eq:rough_bound_32} and \citeeq{eq:rough_bound_94} are perfectly recovered in the MCMC analysis, more strikingly the linear dependence of the limit on $\mchi$. An important fact that cannot be seen from the plot but that we want to emphasize, is the very strong dependence of the limit on the fraction of BH in dark matter, and on the effective deposition efficiency $f_{\rm eff}$. Like in the previous subsection, we can encapsulate our MCMC result in a rough and quick formula that derives from \citeeq{eq:rough_bound_94} (9/4-slope, heavy BH mass regime):
\ben
\sigvav_{\rm max}\approx 10^{-30}{\rm cm^3/s}
\left(\dfrac{\feff\times\fbh}{0.15\times 10^{-6} }\right)^{-3}
\left(\dfrac{\mchi}{100\,{\rm GeV}}\right)\,.
\label{eq:approx_sigvmax}
\een
We emphasize that this limit is valid as long as the BH mass is larger than the breaking mass given in \citeeq{eq:mbreak}, and is in this case fully independent of the BH mass.

\subsection{A tentative PBH interpretation of the HSC microlensing events}
\label{ssec:hsc}
\begin{figure}[t!]
\centering
\includegraphics[width=0.49\linewidth]{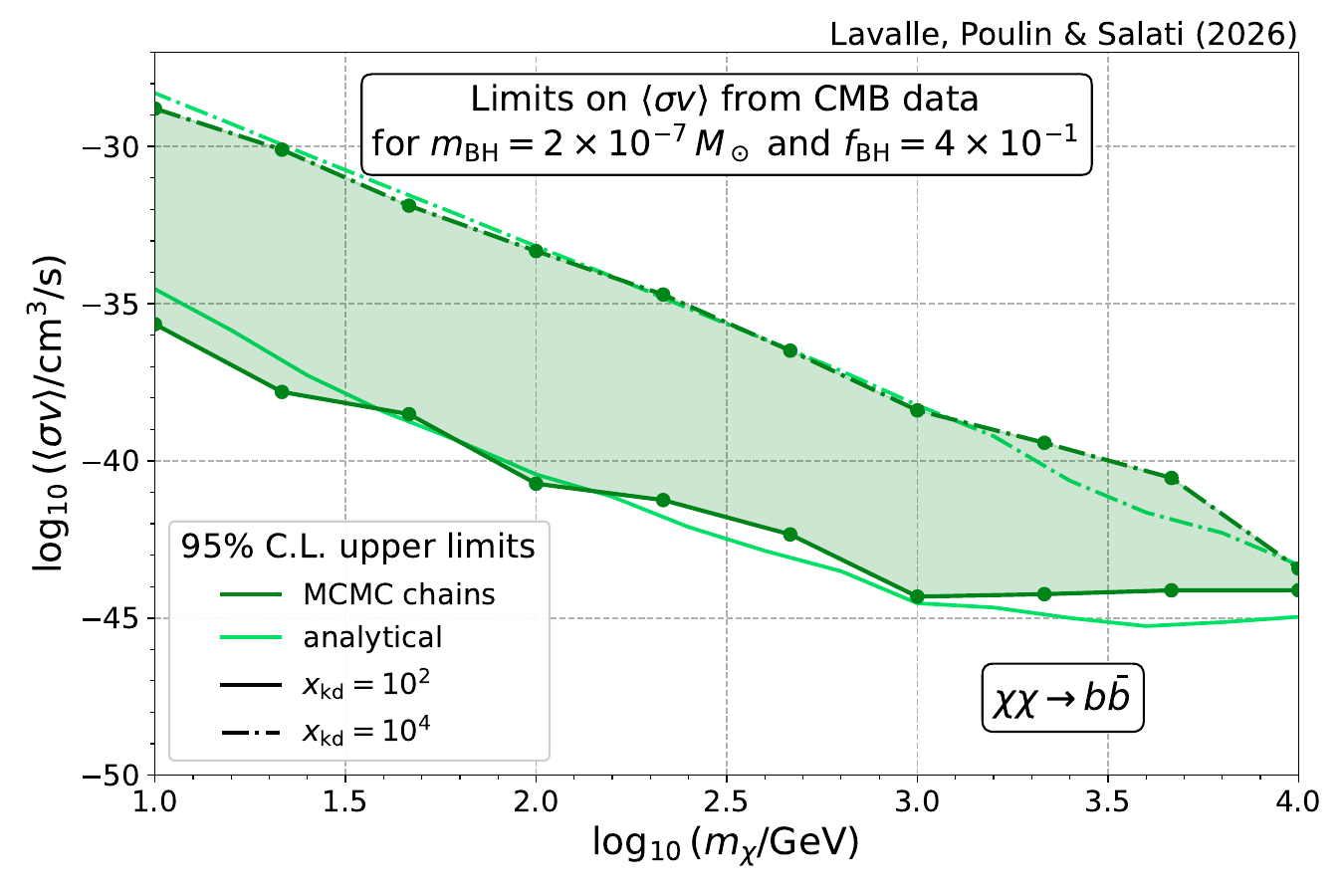} 
\includegraphics[width=0.49\linewidth]{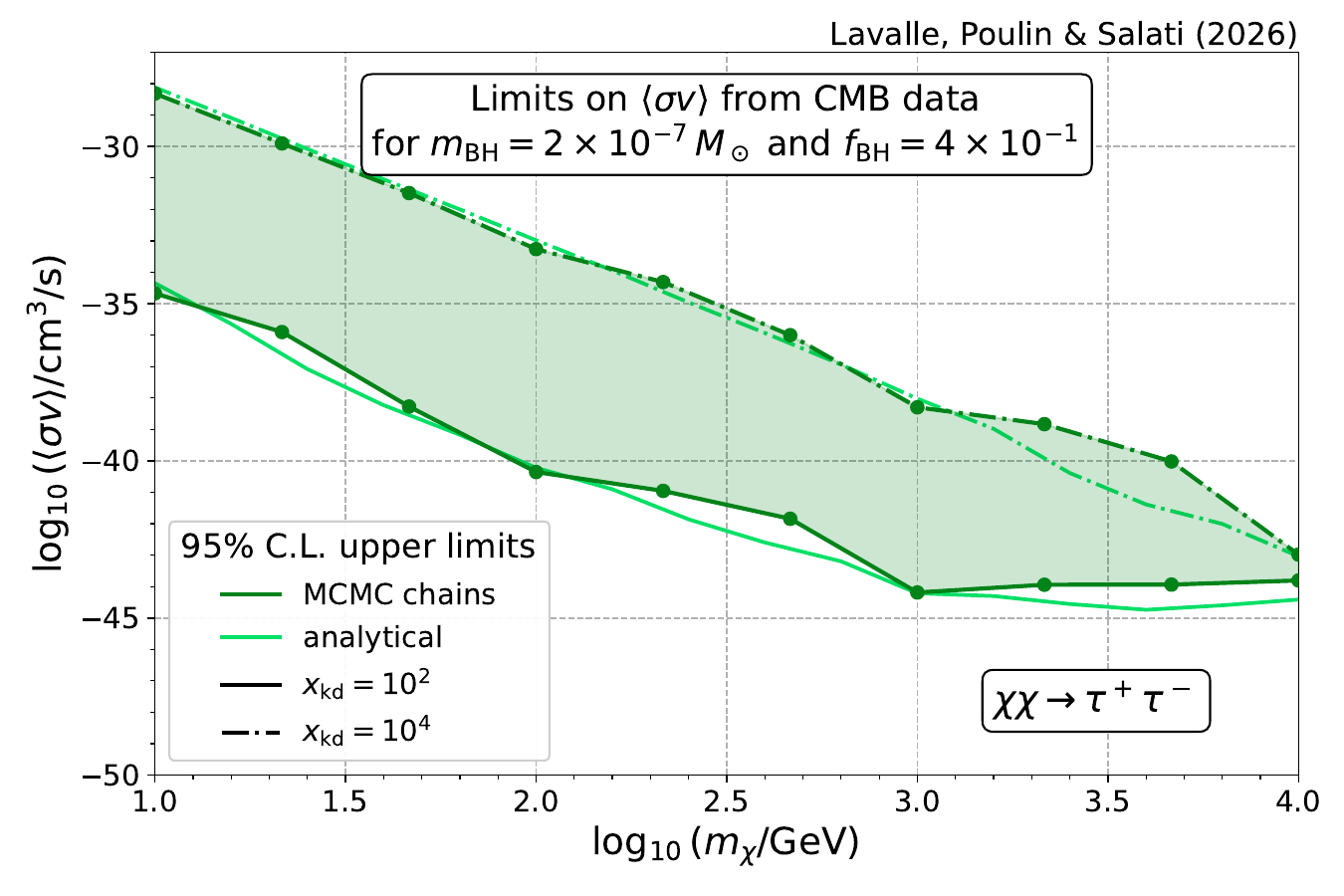} 
\includegraphics[width=0.49\linewidth]{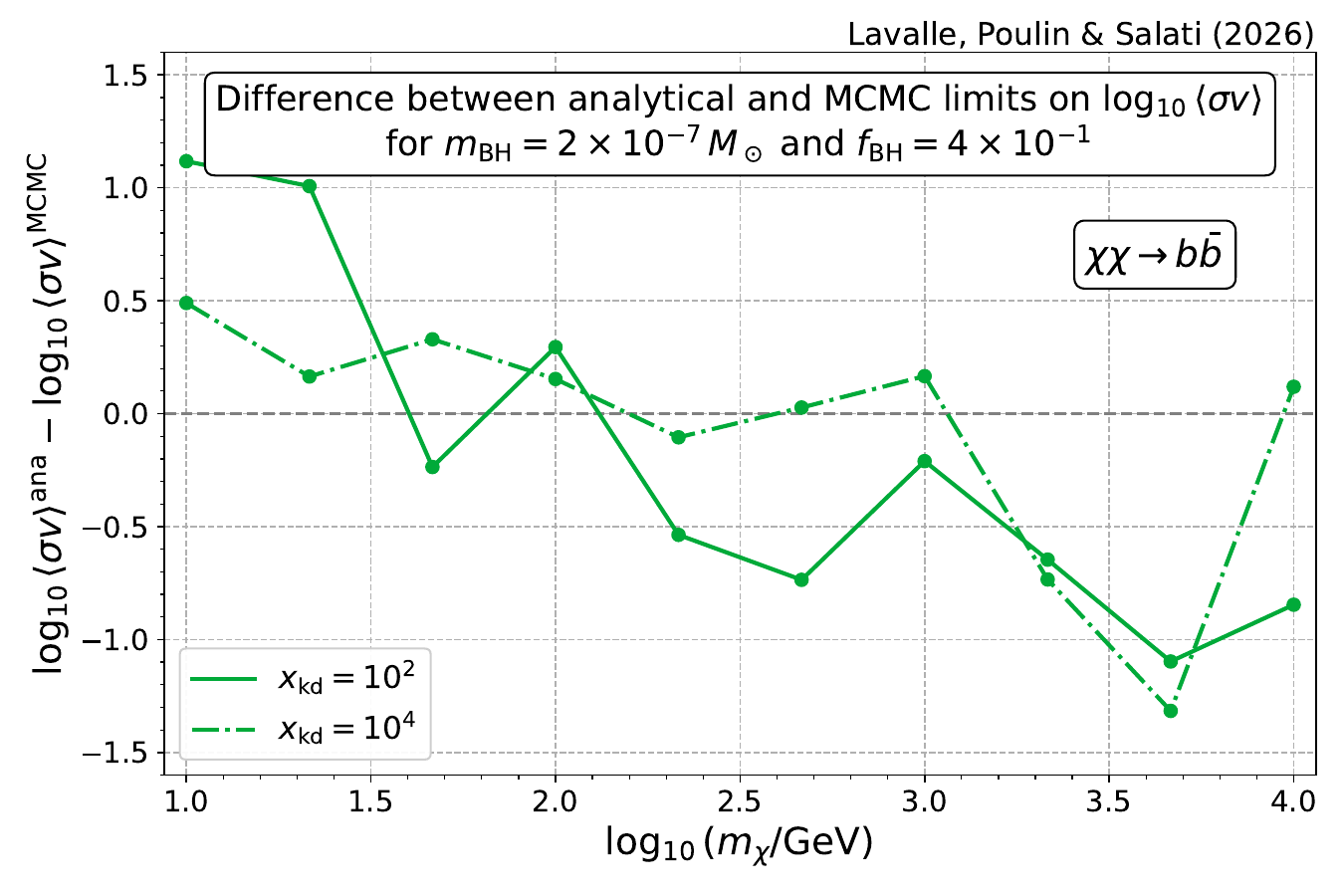} 
\includegraphics[width=0.49\linewidth]{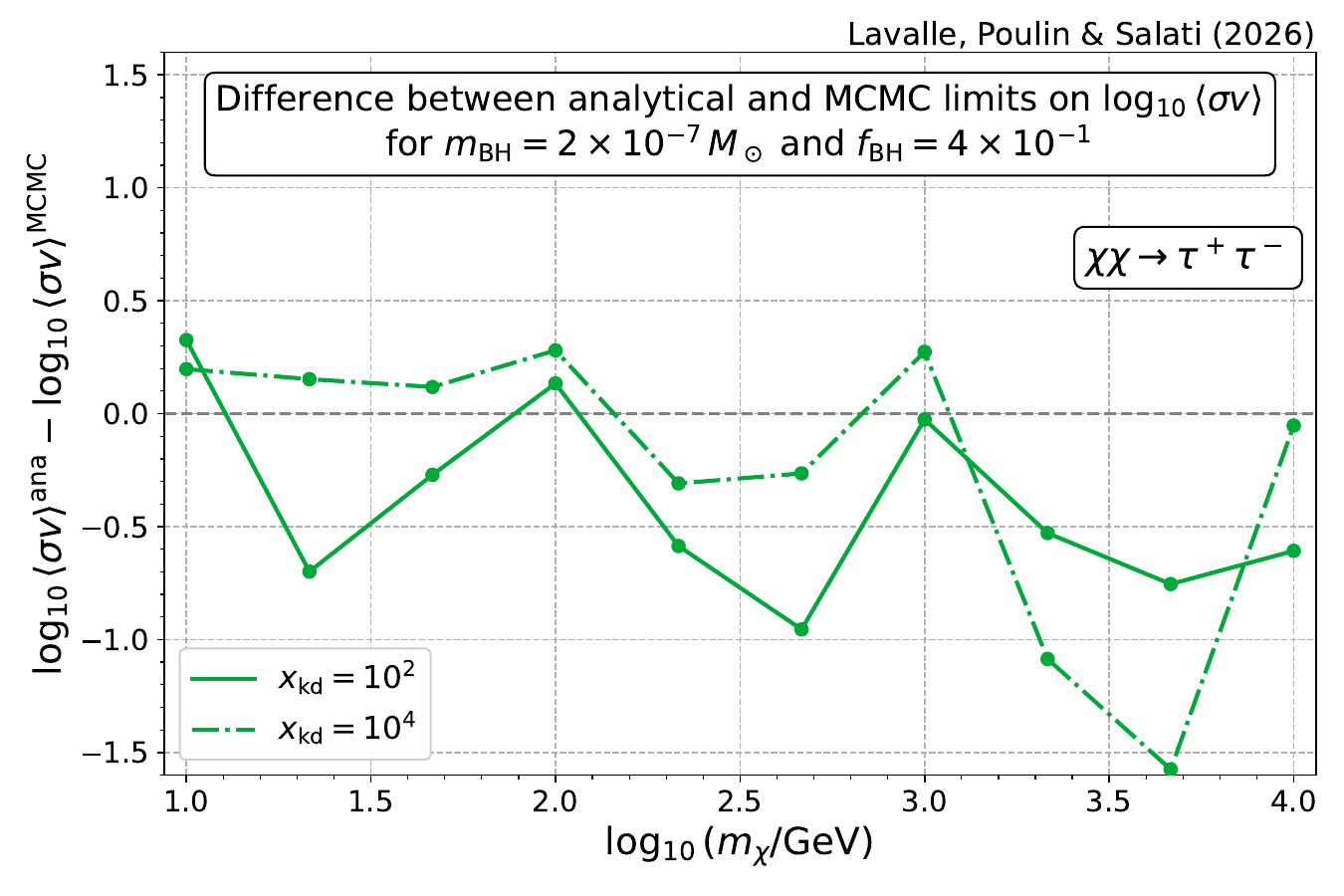} 
\includegraphics[width=0.49\linewidth,valign=c]{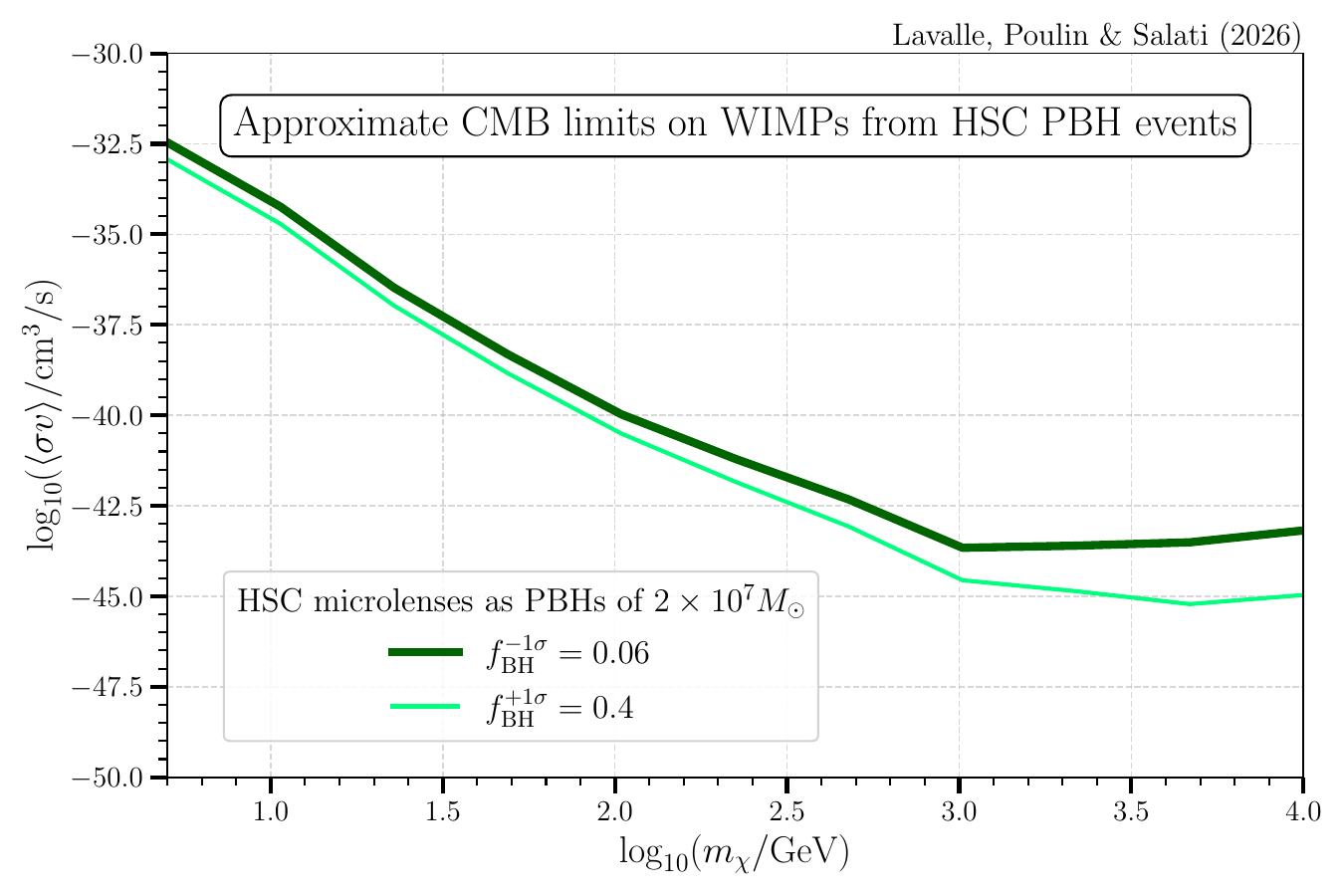}
\begin{minipage}{\dimexpr 0.5\textwidth-\columnsep}
\caption{{\bf Top panels}: Full CMB+BAO+SN1a upper limits obtained on the DM annihilation cross section $\sigvav$ as a function of WIMP mass, assuming the HSC microlenses are due to PBHs (we take the central value for the PBH mass, $2\times 10^{-7}\,\Msun$ and the 1-$\sigma$ upper value of the fraction, $0.4$, inferred from ref.~\cite{SugiyamaEtAl2026}). The green bands show how the limits change when varying $\xkd$ from $10^2$ to $10^4$. Left panel: annihilation into $b\bar b$. Right panel: annihilation into $\tau^+\tau^-$. {\bf Middle panels:} Corresponding differences between the predictions tuned to the best matching $\feff$'s and the MCMC results.
{\bf Bottom panel}: approximate semi-analytical $\pm 1\sigma$ upper limit, using $\feff=0.15$ and $\xkd=100$.}
\label{fig:cmb_limits_on_sigv_hsc}
\end{minipage}
\end{figure}

In this short part, \change{we explore the impact of the recent claim} of the authors or ref.~\cite{SugiyamaEtAl2026} which states that microlensing events observed toward Andromeda (M31) with the Hyper-Suprime-Cam (HSC) instrument onboard the Subaru satellite are consistent with a population of PBHs of mass $\sim 2\times 10^{-7}\,\Msun$ making up a fraction of $0.06-0.4$ ($\pm 1\sigma$ range) of the DM. \change{Although disputed in ref.~\cite{MrozEtAl2026}, it provides an interesting proof of concept of the strength of bounds that could be derived on WIMPs with current sensitivity to PBHs through microlensing.}

In \citefig{fig:cmb_limits_on_sigv_hsc}, we report our MCMC results that we compare with semi-analytical predictions, using the same prescriptions on $\feff$ as in the previous subsection. Results for the $b\bar{b}$ ($\tau^+\tau^-$)  annihilation channel are shown in the left (right) panel. The shaded regions between curves encode the impact of varying the kinetic decoupling time. We have fixed $\mbh = 2\times 10^{-7}\,\Msun$ and $\fbh=0.4$, which corresponds to the upper corner of the preferred region of parameter space in ref.~\cite{SugiyamaEtAl2026}. Interestingly, contrary to the previous cases considered above, this time we mostly lie in the 3/2-slope, light BH mass regime, where the scaling in $\mchi$ radically differs from the 9/4-slope regime discussed above. Indeed, instead of $\sigvav_{\rm max}\propto \mchi$, we now have $\sigvav_{\rm max}\propto \mchi^{-5}$--- see \citeeq{eq:rough_bound_32}, to be compared with \citeeq{eq:rough_bound_94}. Actually, we are quite close to the $3/2\to 9/4$ transition, which is visible at high WIMP masses $\gtrsim 1$~TeV. We see that adopting the tuning of $\feff$ extracted from the previous part allows us to recover a reasonable agreement at least in scaling between our simplified predictions and the full MCMC results --- see the middle panels. We also see that the limits reach a level of cross section that could barely be imagined in standard indirect WIMP searches, down to $\sim 10^{-45}\,{\rm cm^3/s}$ for heavy WIMPs --- in the bottom panel, we show the semi-analytical versions of the upper limits to illustrate the $\pm 1\sigma$ range around the central value of $\fbh$ found in ref.~\cite{SugiyamaEtAl2026} (using $\feff=0.15$ and $\xkd=100$). Therefore, should the interpretation of the HSC events in terms of PBHs be confirmed, that would indeed have serious implication for the presence of WIMPs/FIMPs in our universe.


\section{Summary and conclusion}
\label{sec:concl}
In this paper, we have considered a mixed scenario in which PBHs and WIMPs would co-exist as different DM species. We have assumed a monochromatic mass distribution for PBHs, and $s$-wave annihilation processes for WIMPs. Our model can be fully described with only five parameters: the fraction of DM in PBHs $\fbh$, the PBH mass $\mbh$, the WIMP mass $\mchi$, its annihilation cross section $\sigvav$, and its kinetic decoupling parameters $\xkd$. The general methodology to study the formation of spikes of kinetically decoupled particle DM around PBHs was introduced in ref.~\cite{Eroshenko2016}, which we followed and improved in \paperI\ and \paperII. In the present paper, we have derived limits on such a scenario from a complete statistical analysis of CMB data. These limits are due to the excess of energy deposition in the ambient intergalactic medium following continuous energy injection from the enhanced annihilation rate of WIMPs trapped inside those spikes, which ends up distorting the CMB photon spectrum during and after recombination.

We have made a phenomenological description of the physics at stake in \citesec{sec:recipes}, which can be viewed as complementary to the more technical and general discussion  conducted in \paperII\, where analytical predictions for annihilation signals were obtained in asymptotic regimes and combined together to cover the full range of possibilities. Here, a much more phenomenological approach still allowed us to extract explicit scaling relations in terms of the five model parameters, very useful to understand the final results on qualitative and pedagogical bases (consistent with the more elaborate analytical results). We have first extracted very approximated limits in \citesec{ssec:cmb_approx}, based on several simplifications, like for instance encoding the complex energy deposition history in a mere fudge factor, $\feff=0.1$. These approximate limits indicate that it is difficult for PBHs with masses greater than $\sim 10^{-13}\,\Msun$ to co-exist with WIMPs unless their relative abundance is suppressed as much as $\fbh\lesssim 10^{-7}$. In the same vein, these limits can be turned the other way around, such that an abundance as low as $\fbh=10^{-5}$ of BHs of $10^{-2}\,\Msun$, which could lead to subsolar merger events accessible to future GW experiments, would \eg\ prohibit 1~TeV WIMPs with $s$-wave cross section larger than $\sim 10^{-31}\,{\rm cm^3/s}$.

Then, in \citesec{sec:cmb}, we performed the full statistical analysis of the CMB data by means of the dedicated CLASS code \cite{Lesgourgues2011,BlasEtAl2011} and its ExoCLASS branch \cite{StoeckerEtAl2018}, further backed by the MontePython software \cite{AudrenEtAl2013}, which allowed us to much more carefully account for the complex history of energy deposition pre- and post-recombination. This complete MCMC analysis confirmed the order of magnitude estimates and scaling relations derived from our pedagogical phenomenological approach. It also allowed us to tune the effective deposition parameters so as to get a very good matching between the full MCMC results and semi-analytical predictions condensing the full deposition history into a single fudge factor $\feff$, still much more precise than mere phenomenological estimates. Even if we found values of $\feff$ close to what is expected from analyses of standard WIMP annihilation (no PBHs -- see~\eg~\cite{SlatyerEtAl2009,Slatyer2016}), that would have been difficult to guess so in the first place. Indeed, heating, ionization, and excitation of the intergalactic medium induced by energy cascades, are phenomena controlled by thresholds, and since the coupled PBH-WIMP scenario comes with a time-dependence of energy injection that differs from standard annihilation, we could have expected these effective recipes to lead to significant errors. This is not the case in practice, as shown by our analysis, even though we had to tune the effective efficiency $\feff$ a posteriori to better match the MCMC results (still close to values used in the standard WIMP annihilation case) --- in particular, limits on $\sigvav$ are very sensitive because they scale like $\propto \feff^{-3}$, in contrast with the linear inverse scaling in standard DM annihilation. We provide approximate expressions for these limits, either when assuming that WIMP properties are known [see \citeeq{eq:approx_fbhmax}] or that PBH properties are known [see \citeeq{eq:approx_sigvmax}].

The strength of the obtained limits were expected from earlier work \cite{LackiEtAl2010}, but in this paper, we still improved on two front: (i) the exact semi-analytical determination of the spike profiles backed by a detailed analytical understanding, and devoid of the small mistakes that propagated in the literature and that were pointed out in \paperI\ and \paperII; (ii) a complete CMB data analysis with top-of-the art dedicated tools. On the data analysis aspects, similar efforts were done in ref.~\cite{GinesEtAl2022}, but based on approximations for the spike profiles and related annihilation signals -- our results are in qualitative agreement with theirs, though significantly more precise.

Our CMB limits are complementary to other limits extracted from measurements of the extragalactic diffuse gamma-ray background \cite{BoucennaEtAl2018,CarrEtAl2021c,ChandaEtAl2022,GinesEtAl2022}, though some of these works have inherited part of the small defects mentioned above (see \paperII). \new{As shown in \paperII, a naive and simplified extrapolation of previous gamma-ray data analyses would give more stringent bounds, but a more careful study is required to confirm this trend. Such a work is ongoing, which will soon allow us to check more quantitatively the hierarchy between the CMB and gamma-ray constraints, including the related theoretical uncertainties.} We also leave a thorough study of the effect of velocity-dependent annihilation cross sections ($p$-, $d$-, ..., wave) to future work.

To conclude, it is interesting to notice that the co-existence of PBHs and $s$-wave self-annihilating WIMPs is actually not strictly forbidden. In fact, if PBHs are as light as $\sim 10^{-15}\,\Msun$, corresponding to the hot-spot asteroid mass range, they could live quite peacefully together with WIMPs lighter than $\sim 1$~TeV, or even slightly heavier in the case of late kinetic decoupling. This is somewhat an amusing and exciting situation. Indeed, on the one hand, the 100~GeV - 10 TeV WIMP mass range is still almost free of constraints, as far as indirect detection is concerned \cite{ArcadiEtAl2025,CirelliEtAl2024} -- for WIMPs self-annihilating through $s$-wave processes, \eg\ mediated by pseudo-scalars, direct detection would be essentially blind. Therefore, the detection of WIMPs in this mass range would exclude a sizable fraction of PBHs (say more than $\sim 10^{-7}$), but down to the asteroid mass range only.\footnote{We did not consider extra effects from the population of ultra-compact mini-halos which should also form in this mixed scenario, which should add up significant annihilation signals even with a smaller PBH fraction \cite{ScottEtAl2009b,BringmannEtAl2012b,AbellanEtAl2023}.} On the other hand, the detection of a single subsolar merger event in GWs, a smoking gun for PBHs (even if a tiny DM fraction), would clearly jeopardize the presence of $s$-wave WIMPs in all of the relevant mass range, restricting the annihilation cross section to values $\lesssim 10^{-30}\,{\rm cm^3/s}\,(\mchi/100\,{\rm GeV})\,(\fbh/10^{-6})^{-3}$, even potentially touching the feeble interaction domain (FIMPs \cite{McDonald2002,HallEtAl2010,ChandaEtAl2025}). The tentative interpretation of the HSC microlensing events in terms of PBHs \cite{SugiyamaEtAl2026,MrozEtAl2026} is particularly striking, as shown in \citesec{ssec:hsc}.

\acknowledgments
We thank Pearl Sandick and Tracy Slatyer for insightful discussions on aspects related to this work. JL is grateful to the organizers and participants of the ``No Stone Unturned" workshop held at the University of Utah just before Easter 2025, in particular to Mustafa Amin and Yue Zhao, for interesting conversations on the physics described in this paper. JL is also grateful to the MIT where parts of this study were completed, as complementary to an ongoing MIT/MISTI project. VP is supported by the European Research Council (ERC) under the European Union’s HORIZON-ERC-2022 (grant agreement N$^{\rm o}$ 101076865). This work has benefited financial support from the ANR project ANR-18-CE31-0006 ({\em GaDaMa}), and from European Union’s Horizon Europe research and innovation program under the Marie Sk\l{}odowska-Curie grant agreement N$^{\rm o}$ 101086085–ASYMMETRY---in addition to recurrent public funding from CNRS, the University of Montpellier, and the University of Savoie-Mont-Blanc.

\appendix
\section{MCMC correlations and posteriors}
\label{app:app1}
In this appendix section, as examples, we show the posterior distributions for all the MCMC free parameters in the case the DM fraction in BHs $\fbh$ is let free, while the BH mass and the WIMP parameters are fixed. \citefig{fig:post_fbh_18} illustrates the case of BHs of mass $\mbh=10^{-18}\Msun$ co-existing with WIMPs of mass $\mchi=10$~TeV self-annihilating with the canonical cross section into $b\bar b$ or  $\tau^+\tau^-$, experiencing either early or late kinetic decoupling. \citefig{fig:post_fbh_4} considers heavier BHs with mass $\mbh=10^{-4}\Msun$.

\begin{figure}[t!]
\centering
\includegraphics[width=0.99\linewidth]{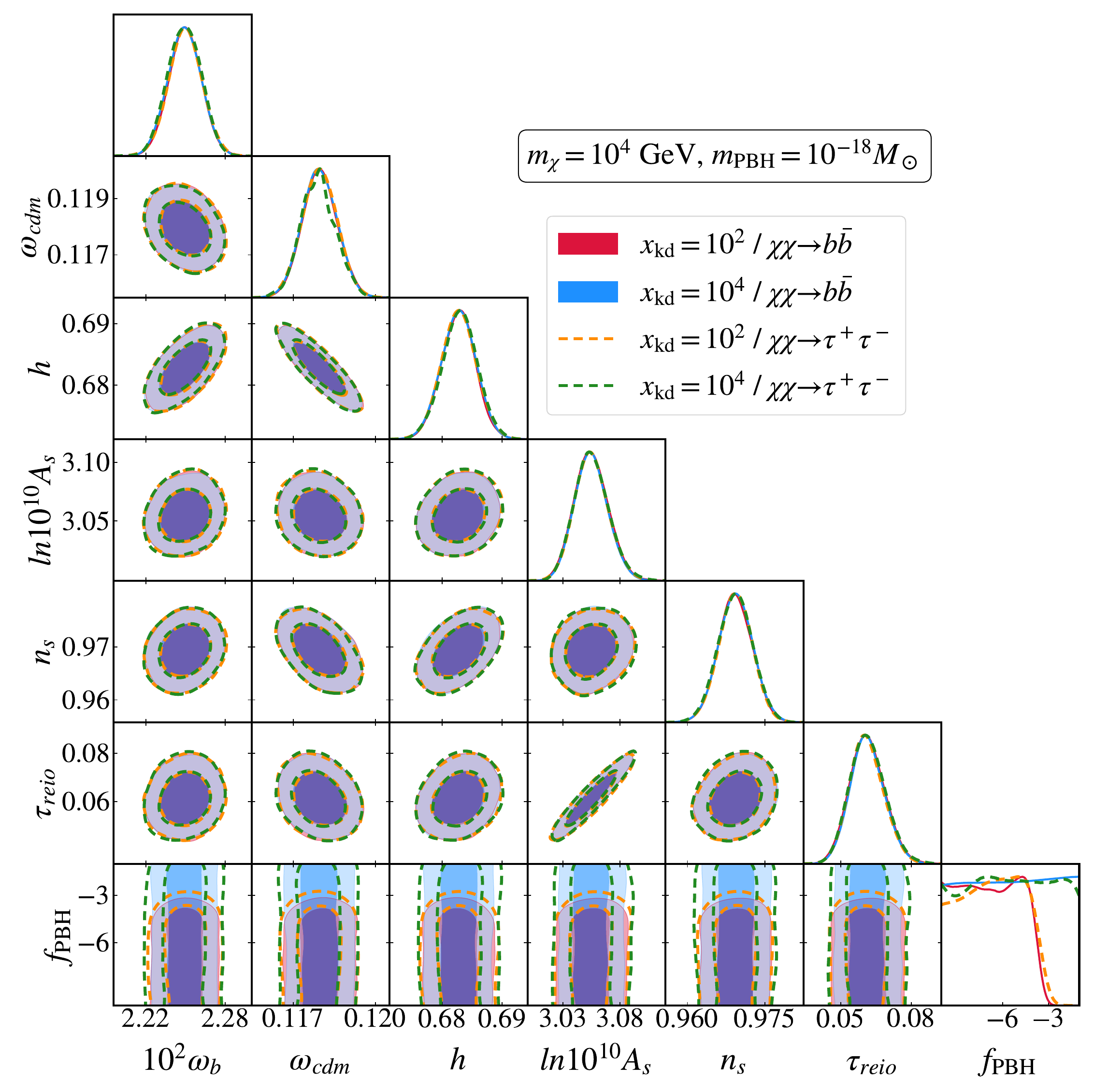} 
\caption{Posterior distributions of the MCMC parameters and their correlations, in the case the DM fraction in BHs $\fbh$ is let free. This example is for BHs of $10^{-18}\Msun$.}
\label{fig:post_fbh_18}
\end{figure}

\begin{figure}[t!]
\centering
\includegraphics[width=0.99\linewidth]{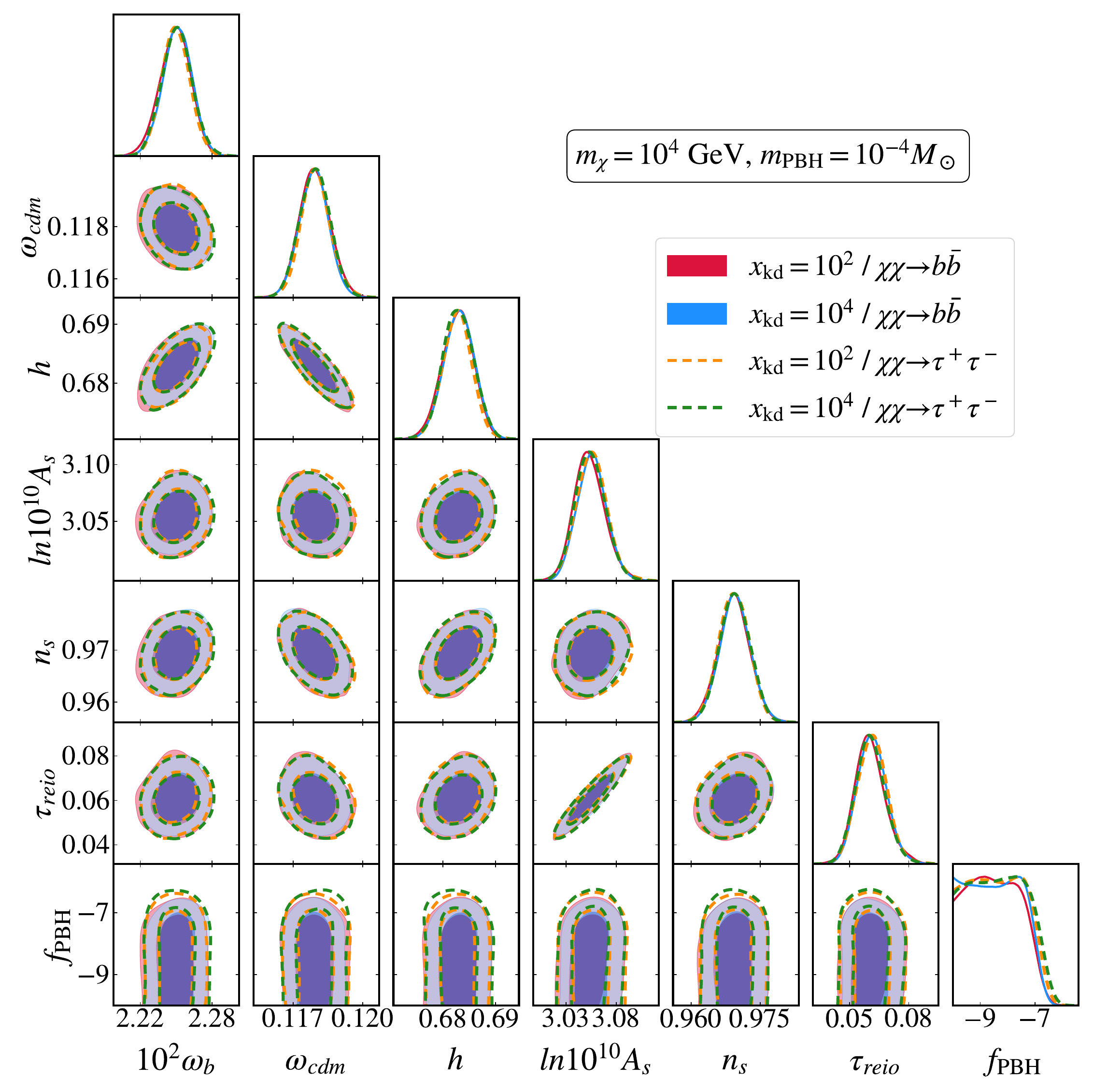} 
\caption{Posterior distributions of the MCMC parameters and their correlations, in the case the DM fraction in BHs $\fbh$ is let free. This example is for BHs of $10^{-4}\Msun$.}
\label{fig:post_fbh_4}
\end{figure}

\bibliographystyle{JHEP.bst}
\bibliography{biblio_dm_spikes.bib}

\end{document}